\documentclass[pra,aps,twocolumn,superscriptaddress,longbibliography]{revtex4-1}
\usepackage[colorlinks=true, citecolor=red, urlcolor=blue ]{hyperref}
\usepackage{graphicx}
\usepackage{bm}
\usepackage{amsmath,amsfonts}
\usepackage{amsthm}
\usepackage{hyperref}
\usepackage{xcolor}
\hypersetup{
    urlcolor=blue,
    citecolor=blue
           }

\usepackage{derivative}
\usepackage{braket}
\theoremstyle{plain}
\usepackage{color}
\usepackage{amssymb}
\usepackage{amsthm}
\usepackage{amsfonts}
\usepackage{float}
\usepackage{tabularx}
\usepackage{graphicx}
\usepackage[utf8]{inputenc}

\usepackage{mathtools}
\usepackage{esvect}
\usepackage{wrapfig}
\usepackage{amsthm}
\usepackage{verbatim}
\usepackage{bbm}
\usepackage[normalem]{ulem}

\usepackage{enumitem}
\usepackage{fmtcount}
\usepackage{booktabs}
\usepackage{csquotes}
\usepackage{epsfig}

\usepackage{tabularx}
\usepackage{graphicx}
\usepackage{amsmath}
\usepackage{braket}
\usepackage{latexsym}
\usepackage{bm}
\usepackage{graphics,epstopdf}
\usepackage{enumitem}
\usepackage{fmtcount}
\usepackage{booktabs}
\usepackage{csquotes}
\usepackage{epsfig}

\theoremstyle{plain}

\def\bea{\begin{eqnarray}}
\def\eea{\end{eqnarray}}
\def\ba{\begin{array}}
\def\ea{\end{array}}

\def\beq{\begin{equation}}
\def\eeq{\end{equation}}

\usepackage[normalem]{ulem}
\usepackage{float}
\usepackage{graphicx}  
\usepackage{dcolumn}          
\usepackage{amssymb}
\usepackage{appendix}
\usepackage{physics}   
\usepackage{mathtools}
\usepackage{esvect}
\usepackage{wrapfig}
\usepackage{amsthm}
\usepackage{verbatim}
\usepackage{bbm}

\usepackage[mathscr]{euscript}

\def\({\left(}
\def\){\right)}
\def\[{\left[}
\def\]{\right]}

\usepackage{orcidlink}

\newtheorem{theorem}{Theorem}

\begin{document}

\title{Positive and non-positive measurements in energy {distillation} from quantum batteries}

\author{Paranjoy Chaki\,\orcidlink{0009-0000-5693-5516}}
\affiliation{Harish-Chandra Research Institute,  
Chhatnag Road, Jhunsi, Prayagraj 211019, India}

\affiliation{Homi Bhabha National Institute,  Training School Complex, Anushakti Nagar, Mumbai 400 094, India}
\author{Aparajita Bhattacharyya,\orcidlink{0000-0002-0090-8735}}
\affiliation{Harish-Chandra Research Institute,  
Chhatnag Road, Jhunsi, Prayagraj 211019, India}

 \affiliation{Homi Bhabha National Institute,  Training School Complex, Anushakti Nagar, Mumbai 400 094, India}
\author{Kornikar Sen\,\orcidlink{0000-0002-7007-0843}}

\affiliation{Harish-Chandra Research Institute,  
Chhatnag Road, Jhunsi, Prayagraj 211019, India}

\affiliation{Homi Bhabha National Institute,  Training School Complex, Anushakti Nagar, Mumbai 400 094, India}

 \affiliation{Departamento de Física Teórica, Universidad Complutense de Madrid, Plaza de las Ciencias 1. Ciudad Universitaria, Madrid 28040, Madrid, Spain}
\author{Ujjwal Sen\,\orcidlink{0000-0002-0091-5847}}
\affiliation{Harish-Chandra Research Institute,  
Chhatnag Road, Jhunsi, Prayagraj 211019, India}
 \affiliation{Homi Bhabha National Institute,  Training School Complex, Anushakti Nagar, Mumbai 400 094, India}

\begin{abstract}
{We investigate energy distillation from quantum batteries within the framework of generalized quantum measurements, including both positive operator-valued measurements (POVMs) and physically realizable non-positive operator-valued measurements (NPOVMs) performed on an auxiliary system coupled to the battery. Two classes of NPOVMs, namely type-1 and type-2, are analyzed in the presence of environmental noise acting on the auxiliary system. We derive general expressions for the distillable energy corresponding to positive and non-positive measurements and show that the distillable energy obtained via NPOVMs remains robust against environmental noise. For a specific model, we demonstrate that the energy extracted using type-1 NPOVMs exceeds or equals that obtained via POVMs under amplitude-damping, dephasing, and bit-flip noise, while type-2 NPOVMs outperform POVMs under amplitude-damping noise. These results establish a clear operational advantage of NPOVMs for energy extraction. We also analyze restricted measurement settings and compare the accessible distillable energy for constrained positive and type-1 non-positive measurements.}
\end{abstract}

\maketitle

\section{introduction}
{Batteries are fundamental energy-storage devices, traditionally based on electrochemical cells that convert chemical energy into electrical energy. The increasing demand for miniaturized technologies has stimulated the development of small-scale energy storage systems, where quantum effects become significant. This has led to the emergence of quantum batteries, which aim to exploit uniquely quantum mechanical features to enhance performance.}

To the best of our knowledge, quantum batteries were first introduced by R. Alicki and M. Fannes in 2013~\cite{Alicki_2013} where the concepts of ergotropy~\cite{Ergotropy,Alicki_2013,batt_rev2} and passive states~\cite{pass1,Pusz_1978,passive3,passive4,passive5,passive6,passive8,ref20,ncptp} were used. The most traditional methods of charging a quantum battery are acting unitary operation directly on the battery~\cite{charg_m,batt_rev2,CQB,charging_3,16} and attaching another system, usually considered as a charger, and operating a unitary on the joint system consisting of the battery and the charger~\cite{charger_1st,Charging_4,batt_rev2,CQB}. The quantities of interest, in these cases, that can be used to gauge the efficiency of a quantum battery are ergotropy, {charging power}, work capacity, etc ~\cite{20,charging_3,charg_m,21,Charging_4}. 
Numerous studies have examined the {charging power} of quantum batteries by considering various models ~\cite{xyz2,cavitya,cavityc,cavityf,26,27,cavityb,cavityd,30,classical,PhysRevA.106.022618,33}. The ultimate bound on work capacity of a quantum battery has been provided in Ref.~\cite{ultimate_bound}. Some works related to open quantum battery is also explored in~\cite{opena,37,opene,openg}. Apart from this, dimensional enhancement in quantum batteries has been studied in~\cite{Ghosh_2022}. To explore the role of quantum properties, such as entanglement and quantum coherence, in the charging and discharging of a quantum battery one can go through Refs.~\cite{Alicki_2013,Entt_1,ent2,PhysRevLett.129.130602,45,ref_455,resource1,Ref_45}.
Experimental studies on quantum batteries can be found in Refs.~\cite{exp3,Hu_2022,ref_52}. A significant number of studies have explored the effect of noise on quantum batteries~\cite{Ghosh_2021,openh,sen2023noisy,56,noise1,st_2,st_3}. 
In Refs.~\cite{demon1,POVM,ord_demon,auxi}, measurement-based methods for charging and discharging energy from quantum batteries have been introduced. 
{Further, the beneficial and detrimental effects of entanglement on the performance of quantum batteries is discussed in Ref.~\cite{Uma_F}.} Apart from that, energy exchange of a dissipative system with a monitor in presence of continuous measurements is given in Ref.~\cite{energy_exchange}.

{In this paper, we introduce non-positive operator-valued measurements (NPOVMs) and employ them for energy distillation from quantum batteries. The distillable energy is defined as the maximum extractable energy obtained by selecting outcomes that yield positive energy, weighted by their probabilities. This notion is motivated by analogous resource-distillation frameworks such as entanglement, coherence, and magic distillation~\cite{D_1,D_2,D_3,D_4,D_5,D_6}. To this end we consider four subsystems: battery, auxiliary, finite dimensional environment, and external. The composite
system of battery and auxiliary undergoes a unitary evolution under the action of a certain Hamiltonian. Henceforth, the auxiliary interacts with the environment and
becomes entangled with it. This interaction can be considered as a noise acting on the auxiliary.
If initially the external is a product with the rest of the system, it remains a product at the end of this process. In such a case, we can perform POVM on the auxiliary by making a projective measurement on the joint
product state of the auxiliary and external system. {On the other hand, an NPOVM can be applied on the auxiliary by performing a projective measurement on the auxiliary-environment (considering the case where the external is initially in product with the rest) or auxiliary-environment-external system (in the situation when the external is initially entangled with the rest), since the auxiliary is entangled with the environment. We refer to these first and second forms of NPOVMs as type-1 and type-2 NPOVMs, respectively. {By NPOVM we mean a measurement that has statistics that cannot be described using positive operators, as is possible with POVMs.} {Importantly, type-1 and type-2 NPOVMs admit a natural experimental realization via global projective measurements on bipartite and multipartite systems, respectively, making our protocol directly relevant for platforms capable of performing global measurements on entangled states. One such experimental work is demonstrated in Ref.~\cite{EXP_NPOVM}, where global jitter measurement is performed on a highly entangled system in an NMR setup.} 
}} {We derive analytical expressions for distillable energy using both positive and non-positive measurements. The maximum stochastically extractable energy follows directly from this by considering the most probable outcome. We also show that the average extractable energy is independent of measurement settings. 
For a fixed battery-auxiliary Hamiltonian, type-1 non-positive measurements yield greater or equal distillable energy under dephasing, amplitude damping, and bit-flip noise. Under amplitude damping, a similar advantage holds for type-2 non-positive measurements over positive ones. 
We also show that the distillable and stochastic extractable energies, obtained using non-positive measurements, are independent of the noise strength.}
{{
{In practical scenarios, implementing the optimal measurement on the auxiliary system can be experimentally demanding, and access to the full set of measurements may be limited. To address this, we restrict the optimization to a physically accessible subset of measurement operators and define the corresponding quantity as accessible energy. This captures energy extraction protocols realizable using only the natural system-environment interaction available in experiments. Under restricted measurements, {POVMs can be advantageous}, notably under dephasing, while NPOVMs remain noise-independent and generally advantageous for bit-flip and amplitude-damping noise.}}} 

{The rest of the paper is structured
as follows: in Sec.~\ref{aA} we describe the concept of positive and non-positive measurements. Sec.~\ref{3S} describes the four subsystems that are involved in energy extraction and how POVM and NPOVM types 1 and type-2 are performed to extract energy.} Sec.~\ref{4f} contains the descriptions of distillable energy, stochastically extractable energy, and average extractable energy. In Sec.~\ref{5S}, we have provided the analytical expression of distillable energy, maximum stochastically extractable energy, and average extractable energy for POVM operations. At the same time, the same task for NPOVM of type-1 and type-2 has been done in Sec.~\ref{6s1} and Sec.~\ref{6s2}, respectively. In Sec.~\ref{7s}, we show that distillable energy and maximum stochastically extractable energy are independent of the type and strength of noise when it is extracted by NPOVM type-1 and NPOVM type-2 operations. 
 {
In Sec.~\ref{8S}, for a fixed battery-auxiliary Hamiltonian and across three noise models, we demonstrate that type-1 NPOVMs outperform POVMs in both energy distillation and maximal stochastic energy extraction.}
Moreover, we have done a comparative study of NPOVM type-2 and POVM in terms of distillable energy and maximum stochastically extractable energy in the presence of amplitude damping noise in Sec.~\ref{11S}. Sec.~\ref{12S} contains the detailed information about the accessible distillable energy and stochastically accessible energy. Finally, in Sec.~\ref{CCCCCCCCCCC}, we present the concluding remarks.
{\section{General POVM and NPOVM}\label{aA}}
{In this section, we first briefly recapitulate positive operator-valued measurements (POVMs) and introduce non-positive operator-valued measurements (NPOVMs).} 

{Let us consider a system, $S_1$, attached to an environment, $S_2$. Let initially the composite state of $S_1$ and $S_2$ are prepared as $\rho_{S_1}\otimes\rho_{S_2}$, i.e., in a product state, where $\rho_{S_1}$ and $\rho_{S_2}$ are states of the system and environment, respectively. If a projective measurement, with projection operators, say $\{\ket{\Psi_i}\bra{\Psi_i}\}_i$, is performed on $\rho_{S_1}\otimes\rho_{S_2}$ and  $\ket{\Psi_i}\bra{\Psi_i}$ gets clicked then the final state of the system, $S_1$, after the measurement would be 
$$\rho^i_{S_1}´=\frac{\tr_{S_2}(\ket{\Psi_i}\bra{\Psi_i}\rho_{S_1}\otimes\rho_{S_2} \ket{\Psi_i}\bra{\Psi_i})}{\tr(\ket{\Psi_i}\bra{\Psi_i}\rho_{S_1}\otimes\rho_{S_2} \ket{\Psi_i}\bra{\Psi_i})}=\frac{\chi_i \rho_{S_1} \chi_i^{\dag}}{\tr( \rho_{S_1} \chi_i^{\dag}\chi_i)}.$$
Here, ${\chi}_i$ serves as the effective measurement operator that acts on the state of the system $S_1$. The measurement induced on the subsystem, $S_1$, by performing a projective measurement on the bigger system, $S_1S_2$, that contains $S_1$, is popularly known as positive operator-valued measurement or in short POVM. The reason behind this nomenclature is that the probability, $p_i$, of getting a particular outcome, say $\rho^i_{S_1}$, can be expressed as $p_i=\tr( \rho_{S_1} \chi_i^{\dag}\chi_i)=\tr( \rho_{S_1} E_i)$, which is a function of the positive semi-definite operator, $E_i=\chi_i^{\dag}\chi_i$. Each operator, $E_i$, is often called POVM element. Any set of POVM elements, $\{E_i\}$, {must meet the following properties:~\cite{nielsen00} }}
{\begin{itemize}
    \item  $E_i$ is Hermitian, i.e., $E_i=E_i^{\dag}$, $\forall$ $i$. 
\item Each of the POVM elements, $E_i$, is positive semi-definite, i.e., it has only non-negative eigenvalues. 
\item The sum of all elements of the set, $\{E_i\}_i$, is equal to the identity operator, i.e.,  $\sum_{i} E_i=\mathbb{I}$, where $\mathbb{I}$ is the identity operator which acts on the Hilbert space of the subsystem, $S_1$.
\end{itemize}
According to {Naimark's dilation theorem \cite{Neimark}}, any set of operators, $\{E_i\}$, acting on a system, $S_1$, that satisfy the above-mentioned properties can be thought of as a set of POVM elements, which can be produced by attaching an external system, $S_2$, to $S_1$ and performing a projective measurement jointly on $S_1$ and $S_2$.}

{The concept of POVMs depends on the tacit assumption that the initial state, $\rho_{S_1}\otimes\rho_{S_2}$, on which projective measurements would be performed, is a product of the individual states of $S_1$ and $S_2$. This motivates us to introduce a more generalized measurement by relaxing this restriction on the initial state of $S_1$ and $S_2$. If we initially prepare the pair of systems, $S_1$ and $S_2$, in a state $\rho_{S_1S_2}$ which may not necessarily be a product or even separable, perform 
the projective measurement, {$\{\ketbra{\Psi_i}\}_i$}, on the entire arrangement, and get the output $\ketbra{\Psi_i}$, then the final state of the system, $S_1$, after the application of the measurement, would be}
{\begin{equation*}
\tilde{\rho}^{i}_{S_1}=\frac{\tr_{S_2}\left(\ketbra{\Psi_i}\rho_{S_1S_2}\ketbra{\Psi_i}\right)}{\bra{\Psi_i}\rho_{S_1S_2}\ket{\Psi_i}}.
\end{equation*}}
{In such a scenario, where the initial state, $\rho_{S_1S_2}$, is not necessarily product, the probability of getting any particular outcome, in general, would not be expressible in terms of positive semidefinite operators. Therefore we refer to the effective measurement performed on $S_1$ by applying joint projective measurement on an entangled state of the system $S_1S_2$ as a non-positive operator-valued measurement or NPOVM.} 

{In Fig.~\ref{xxpo}, we describe the PV, POVM, and NPOVM methods by providing schematic diagrams. In diagram (a), we describe the PV measurement protocol where a measurement on the product system (AB) is performed in the bi-orthogonal product basis. Tracing out B causes an effective projective measurement on A. Instead, if we apply a global unitary to the bi-orthogonal product basis and then measure the system in the rotated basis, the resulting measurement basis no longer stays in the bi-orthogonal product form. Therefore, the resulting measurement on A is no longer a strict projective measurement and becomes a POVM. In the diagram (c), we consider the case where the initial state is entangled. In this case, the measurement in the rotated basis, which, in general, is not biorthogonal, can cause an NPOVM. If the measurement following the protocol described in diagram (c) provides benefits over all the POVM operations, then the measurement is confirmed as an NPOVM.}

\section{Preparation of the system for the measurements}\label{3S}

Here we discuss the system's grooming before the measurements take place for energy extraction from the battery. In addition to the quantum battery ($B$), our setup involves three more bodies: an auxiliary  ($A$), a finite dimensional  environment ($E$), and an external ($X$). Let the initial state of the entire system ($S$) be $\rho^0_{S}=\rho_{BAE}^0\otimes \rho_{X}$.
Let the identity operators acting on the Hilbert spaces that describe $B$, $A$, $E$, and $X$ be, respectively, $\mathbb{I}_B$, $\mathbb{I}_A$, $\mathbb{I}_E$, and $\mathbb{I}_X$. 

The energy extraction protocol involves the application of POVM and NPOVM on the auxiliary system, $A$. In general, initially, $B$ and $A$ may not have any correlation. But for the operations performed on $A$ to affect $B$, we need $B$ and $A$ to be entangled. 
Therefore, we first apply a unitary $U_{BA}$ jointly on $BA$. During this evolution, the environment and the external system act as spectators. Hence the ultimate state of $S$ after the application of $U_{BA}$ is $\rho_{S}^1=(U_{BA}\otimes \mathbb{I}_{E}\otimes \mathbb{I}_{X}) \rho^{0}_S(U_{BA}^\dagger\otimes \mathbb{I}_{E}\otimes \mathbb{I}_{X})=\rho_{BAE}^1\otimes \rho_{X}$, where $\rho_{BAE}^1=(U_{BA}\otimes \mathbb{I}_{E}) \rho^{0}_{BAE}(U_{BA}^\dagger\otimes \mathbb{I}_{E})$. Then a global unitary gate $U_{AE}$ is jointly acted upon $AE$ which transforms $BAE$ to $\rho_{BAE}^2=(\mathbb{I}_B\otimes U_{AE})\rho_{BAE}^1(\mathbb{I}_B\otimes U_{AE}^\dagger)$, and keeps the external still fixed at $\rho_X$. Hence the final state of the system at this moment is $\rho_S^2=\rho_{BAE}^2\otimes \rho_X$. The reason behind this evolution can just be an uncontrollable effect of the environment on the auxiliary or can also be considered as a manually created interaction by the technician to extract energy from the battery. This interaction between $A$ and $E$ will be utilized to perform NPOVM on $A$.  In the next part, we separately discuss in detail the roles of POVM and type-1 and {type-2} NPOVMs performed on auxiliary, $A$, in extracting energy from $B$. The protocols we use for extraction of energy from a battery using POVM and type-1 and type-2 NPOVMs are depicted 
through a schematic diagram (refer to Fig.~\ref{figx3}). 

As the simplest non-trivial example, we choose the battery, auxiliary, environment, and external to be qubits. The local Hamiltonian of the battery (auxiliary/environment) is taken as $H_{B(A/E)}=h_{B(A/E)}\sigma^z$, where $\sigma_z$ is a Pauli matrix. We consider both the battery, $B$, and the auxiliary, $A$, to be initially prepared in the excited states, i.e., respectively, $\ket{e}_B$ and $\ket{e}_A$, whereas the environment, $E$, qubit is taken to be in the ground state, $\ket{g}_E$. Hence the joint state of the system, $BAE$, is given by $\ket{e}_B\bra{e} \otimes \ket{e}_A\bra{e}\otimes \ket{g}_E\bra{g}$.

\begin{figure}
		\centering
			\includegraphics[width=8.5cm]{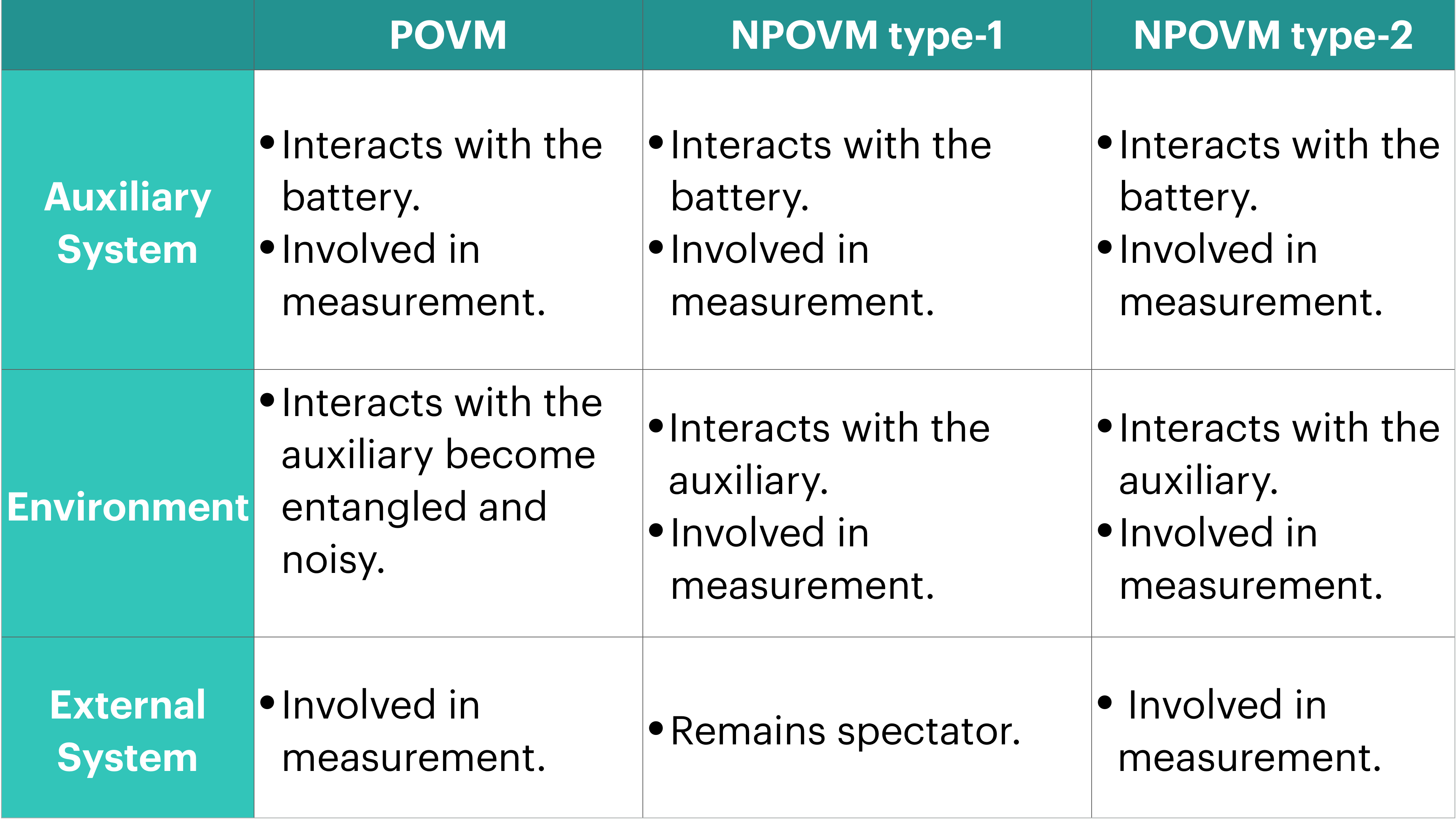}
\caption{Table to mention the role of auxiliary, environment, and external system. }
		\label{ft1}
		\end{figure}

\begin{figure}
		\centering
			\includegraphics[width=8.5cm]{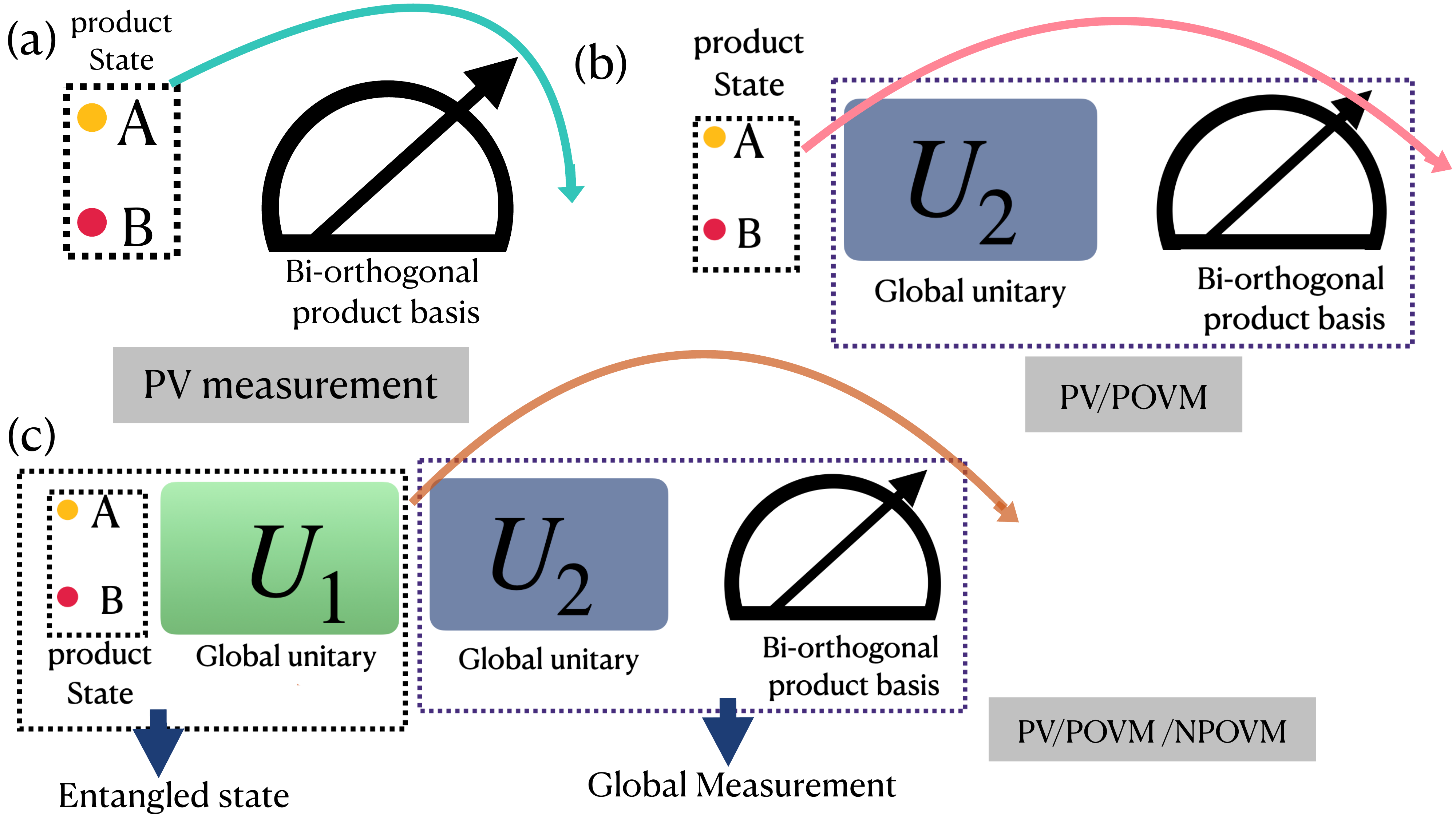}
\caption{{\textbf{Description of general PV, POVM, and NPOVM.} In the diagrams (a), (b), and (c), PV, PV/POVM, and PV/POVM/NPOVM are depicted, respectively. The black arrow in each figure denotes the measurement in the biorthogonal product basis. In diagrams (b) and (c), the unitary $U_2$ rotates the biorthogonal product basis in an arbitrary direction, making the resulting basis of non-biorthogonal form. In diagram (c), we use another unitary $U_1$ to make the initial state of AB entangled. In all the diagrams, A and B systems are denoted by yellow and red dots.}}
		\label{xxpo}
		\end{figure}

{ {In Fig.~\ref{figx3}, we illustrate the protocols of of energy extraction for POVM, NPOVMs of type-1
and type-2 
using schematic diagrams. Two types of schematics are provided: the upper diagram presents the full procedure in detail, while the lower diagram offers a simplified circuit representation for clearer understanding.}
In the left, middle, and right panels, we provide schematic diagrams portraying the process of 
measurement-based extraction of energy by performing, respectively, POVM and NPOVM of type-1 and type-2, on an attached auxiliary. The procedure involves four systems: the battery (orange dot), auxiliary (green dot), environment (dark-green dot), and an external system (violet dot). Each solid elliptical shape represents the interaction taking place between the systems contained in the ellipse. The joint projective measurements being performed on the constituents are depicted using hollow elliptical shapes. The steps involved in the processes are described below in the diagrams. {For better understanding, we also provide a circuit  diagram based on quantum gates for clear understanding of the process. In the left, middle, and right-handed diagrams, POVM, NPOVM of type-1, and NPOVM of type-2 operations on the auxiliary system are described. In all of the cases, we take four systems: battery, auxiliary, environment, and external system. Firstly, the battery and auxiliary system undergo unitary evolution. 
This evolution is described by the blue box representing the unitary gate $U_{BA}$ acts on battery and auxiliary; after that, the auxiliary interacts with the environment by the global unitary evolution, described by the unitary gate ($U_{AE}$), denoted by the green box and becomes }{noisy}. { Now for the leftmost and middle figures, i.e., for POVM and NPOVM of type-1, the external system remains as a product with the rest of the system. }
{
For NPOVM of type-2, we consider the initial entanglement between the external system and the rest of the system. To perform POVM measurement on the auxiliary system, we perform projective measurement on the auxiliary and external system and then focus on the auxiliary system. On the other hand, for NPOVM of type-1, projective measurement is performed on the composite system of auxiliary and environment, after that focusing on the auxiliary. At last, for NPOVM type-2 measurement, the projective measurement is performed on the joint state of auxiliary, environment, and external system, and after that discarding the environment and external system. Usually the basis at which the projective measurement is performed, is realized by implementing unitary gate on the computational basis. The unitaries involved in POVM, NPOVM of type-$1$ and type-$2$ measurements are denoted as $U^P_M$, $U^{N^1}_M$, $U^{N^2}_M$ in violet boxes. On the other hand the local measurements in the computational basis are denoted as the black arrows.}} {For a better understanding of the respective roles of the auxiliary, environment, and external systems - that is, the systems other than the battery - we provide a table in Fig.~\ref{ft1}. }

{The environment can be regarded as an additional system that interacts locally with the auxiliary system, independently of the other subsystems, leading to the formation of entanglement between the environment and auxiliary. The role of the environment is to introduce noise into the auxiliary system, while getting entangled with the auxiliary. During NPOVM operation of type-1, the entanglement between the auxiliary and the environment is treated as the resource. For NPOVM of type-2, the joint initial state of the battery, auxiliary, environment and the external is entangled. In this case, the entanglement between the auxiliary and the environment, as well as the external system, acts as the resource. 
Moreover, the control over the environment is considered in the protocol NPOVM type-1 and type-2 operations. 
 Therefore, this control determines whether we can perform a joint projective measurement on the combined system of auxiliary and environment during the NPOVM operation. Specifically, when the measurement includes the environment, which is entangled with the auxiliary, we can obtain an NPOVM on the auxiliary system, whereas excluding the environment yields a POVM in our protocol and realized by performing a global projective measurement on the auxiliary and external system, which is in a product state.}

{In the process of making a measurement on the auxiliary, we first apply a unitary operation $U_{BA}$, generated by the transverse-field Hamiltonian $H_{BA}$, on the battery and auxiliary systems. Subsequently, a unitary $U_{AE}$ is applied to the auxiliary and environment systems. Conceptually, this process can be assumed as the operation $U_{BA}$ occurs on a timescale much faster than the dynamics induced by other unitaries, like $U_{AE}$ governed by other coupling Hamiltonians.}

{\section{Figures of Merit}\label{4f}}
{{In our measurement-based protocol for energy extraction, there can be measurement outcomes for which the energy difference is comparatively large, but the probability of occurrence of that particular outcome may be very small. To avoid potential misinterpretations, instead of completely focusing on the energy difference, we consider the product of the energy difference and the probability of obtaining that specific outcome and define two figures of merit based on this. }
}{
The definitions of the two figures of merit, namely the distillable energy and the maximum stochastically extractable energy are outlined in this section. The distilled energy for a given measurement setting is defined as the sum of the probabilities of the measurement outcomes multiplied by the corresponding energy differences between the initial state and final measurement outcomes of the battery, which are positive. Now, if we perform an optimization over all possible measurement settings, then the maximum distilled energy is regarded as the distillable energy. Since our focus is on energy extraction within a measurement-based protocol, we consider the outcomes for which the energy difference between initial energy of the battery and those outcomes are positive.} 

{{Let us consider $\Delta E_i$ to be the extractable energy corresponding to the $i^{\text{th}}$ outcome of any given measurement setting with probability $p_i$. 
The distilled energy for that  particular measurement setting, at any given time $t$, is given by $S_d=\sum_{i\in\mathcal{I}}p_i\Delta E_i$. Here, $\mathcal{I}$ is the set of outcomes for which $\Delta E_i$ is positive. 
For instance, let us suppose that the extractable energy corresponding to the $i^{\text{th}}$ outcome is given by  $\Delta E_i=E^{in}-E^f_{i}$, for a given measurement setting.  Here, $E_{in}$ is the initial energy of the battery, and $E_i$ denotes the energy of the final state corresponding to the $i^{\text{th}}$ outcome. Now, suppose $\Delta E_i > 0$, for $i \le 5$, then $\mathcal{I}\in\{1,2,3,4,5\}$.
If we further optimize the distilled energy for all possible set of measurement operators then the maximum is referred to as the distillable energy denoted as $S_D$. At the same time, the stochastically extractable energy corresponding to any given outcome is the energy difference of the initial state and the final outcome times the probability of the outcome, defined as $S^P_i=p_i \Delta E_i$. The maximum value of this stochastically extractable energy is termed as the maximum stochastically extractable energy denoted as $S^P$. 
On the other hand, if we consider the extractable energy over all possible outcomes and take its average, it is referred to as the average extractable energy, denoted by $S_{av}$.
}} \\
\section{Application of POVM on the auxiliary}\label{5S}
{Here, we provide the analytical form of distillable energy, maximum stochastically extractable energy, and average energy using POVM.}

{Let us consider a composite quantum system ($S$) consisting of a battery $(B)$, an auxiliary $(A)$, an environment $(E)$, and an external ($X$). The aim is to extract energy using POVM. 
 Let immediately before the application of the measurements, the states of $BAE$, $X$, and $S$  are $\rho^2_{BAE}$, $\rho_X$, and $\rho^2_S=\rho_{BAE} \otimes \rho_X$, respectively. The battery's energy is defined by the Hamiltonian, $H_B$.}
{We can define the following operators: $\rho_B = \operatorname{tr}_{AE}[\rho_{BAE}^{\,2}]$,
$Z=\text{tr}[\rho_{B} H_B]\mathbb{I}_{BAX}-H_B \otimes \mathbb{I}_{AX}$,
and
$\mathcal{B} = \text{tr}_B\left[\text{tr}_E\left(\rho^2_{S}\right)Z\right]$, where $\mathbb{I}$ denotes the identity operator acting on the relevant subsystem. The set of eigenvalues of $\mathcal{B}$ can be denoted as $\{\beta_i\}_i$.} 

\begin{theorem}
{The distillable, maximum stochastically extractable, and average extractable energies by performing the optimal POVM on the auxiliary are given by
$S^{P}_D = \sum_{i \in \mathcal{I}} \beta_i$, 
$S^{P}_{\max} =\max_i \beta_{\max}$, 
and 
$S^{P}_{av} = \sum_i \beta_i$, respectively,
where $\mathcal{I}$ denotes the set of indices denoting the positive eigenvalues among $\{\beta_i\}$.}
\end{theorem}

 \begin{proof}
For the details on a general POVM one can go through Sec.~\ref{aA}.
Here we discuss the POVM that is applied on $A$ for extraction of energy from $B$. 
 It is evident from the expression of $\rho_S^2$ that, $\rho_S^2$ can be entangled in the bipartition $B:AEX$ and $BA:EX$ but is definitely separable in the bipartition $BAE:X$. In this situation, a projective measurement, $\{\ket{\Psi_i}_{AX}\bra{\Psi_i}\}_i$, is performed on the joint state of $A$ and $X$, and a particular outcome, say $\ket{\Psi_i}_{AX}\bra{\Psi_i}$, is selected. 
Since $AX$ does not share any entanglement, the measurement effectively reduces to a POVM applied on $A$. 
\begin{figure*}
		\centering

\includegraphics[width=15cm]{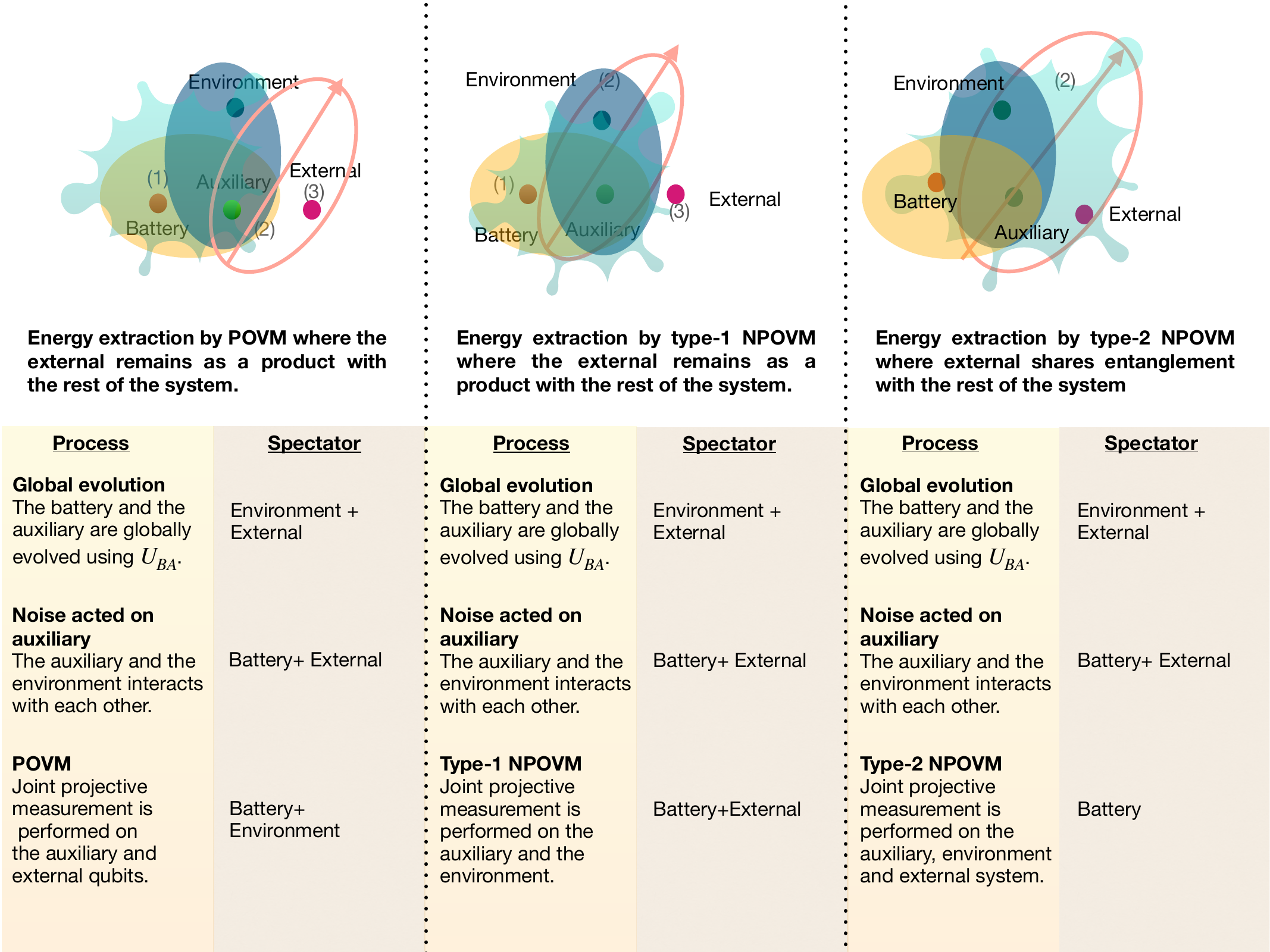}            \includegraphics[width=5.6cm]{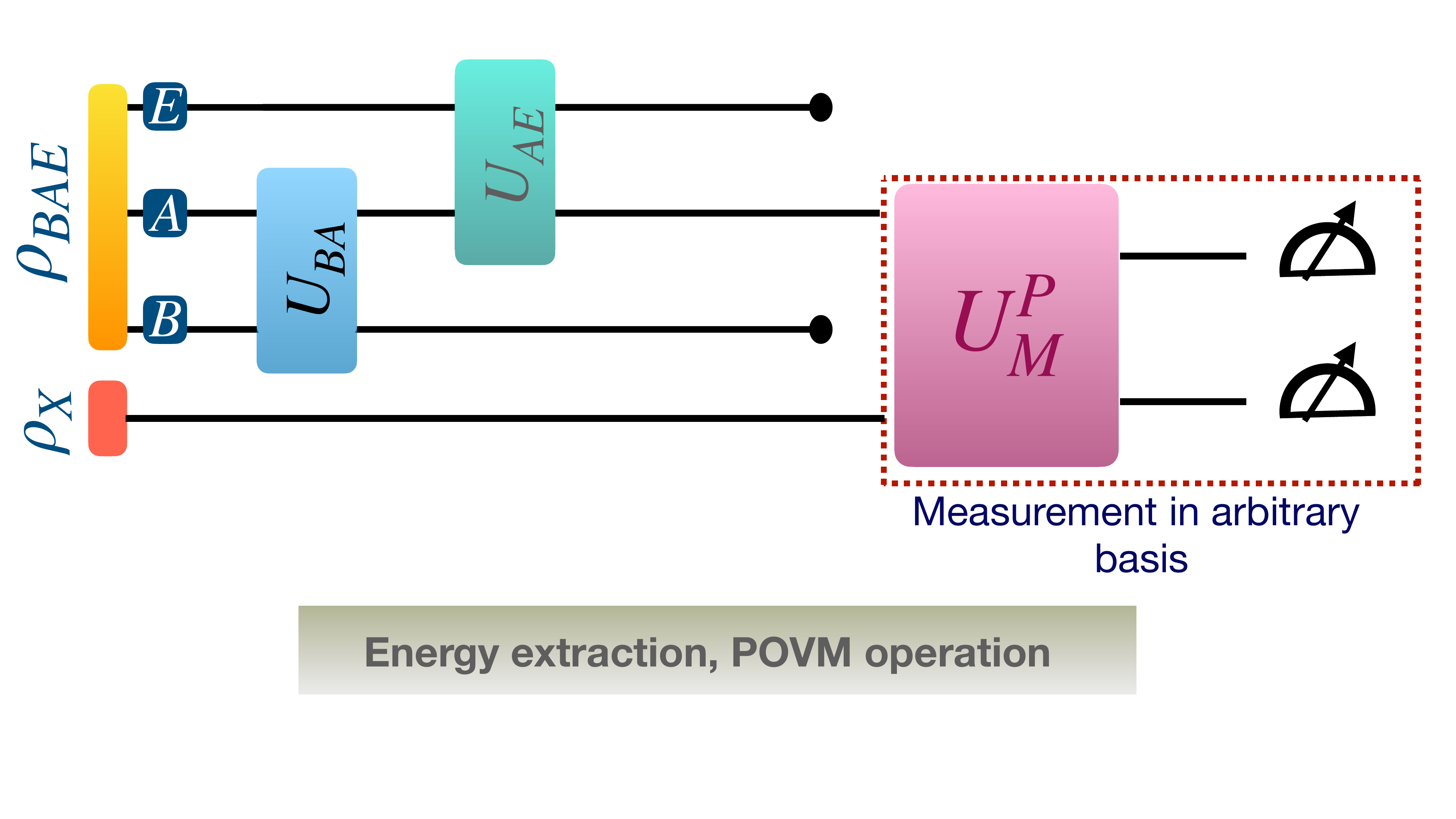}
 \includegraphics[width=5.6cm]{ 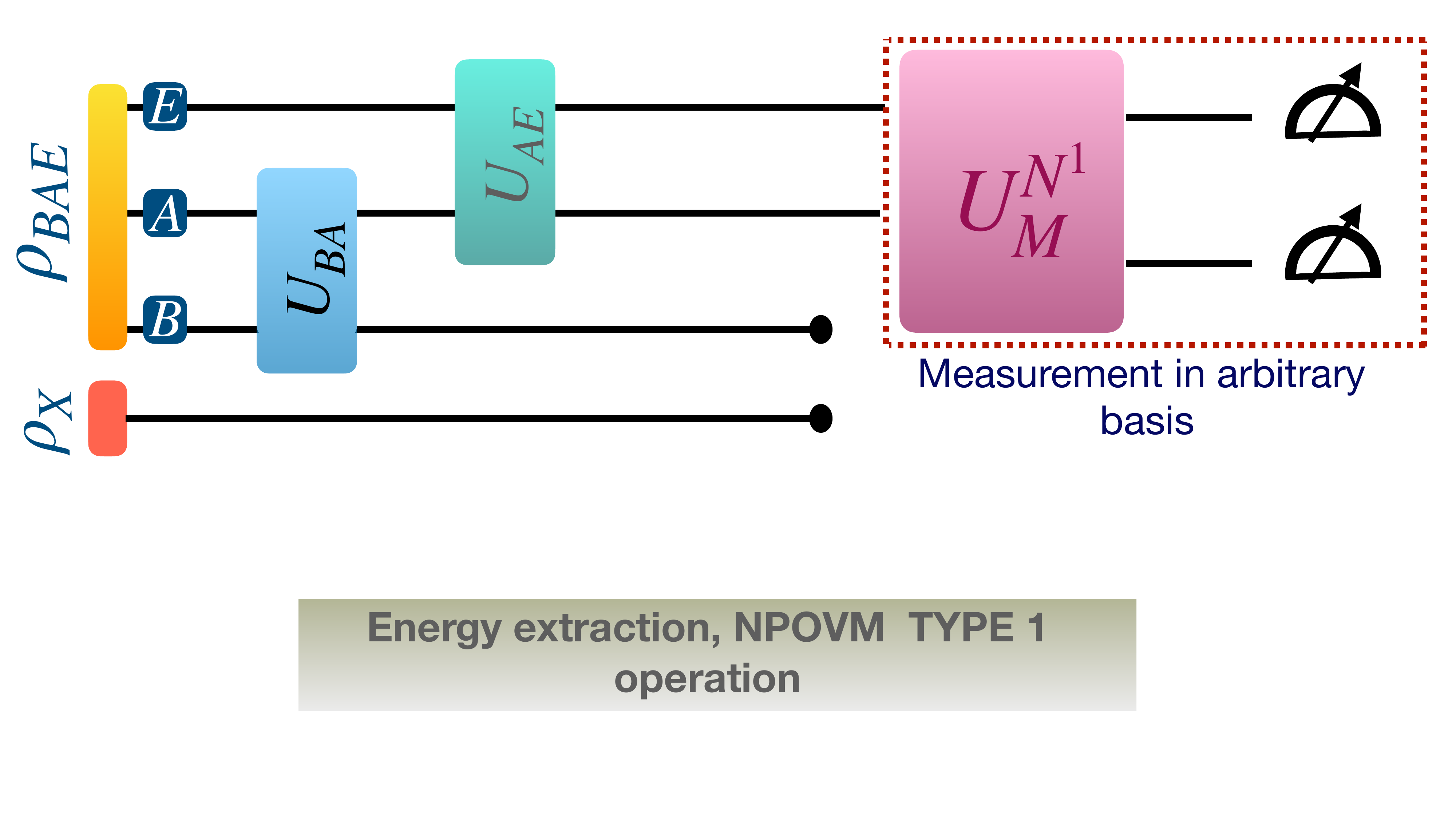  }
   \includegraphics[width=5.6cm]{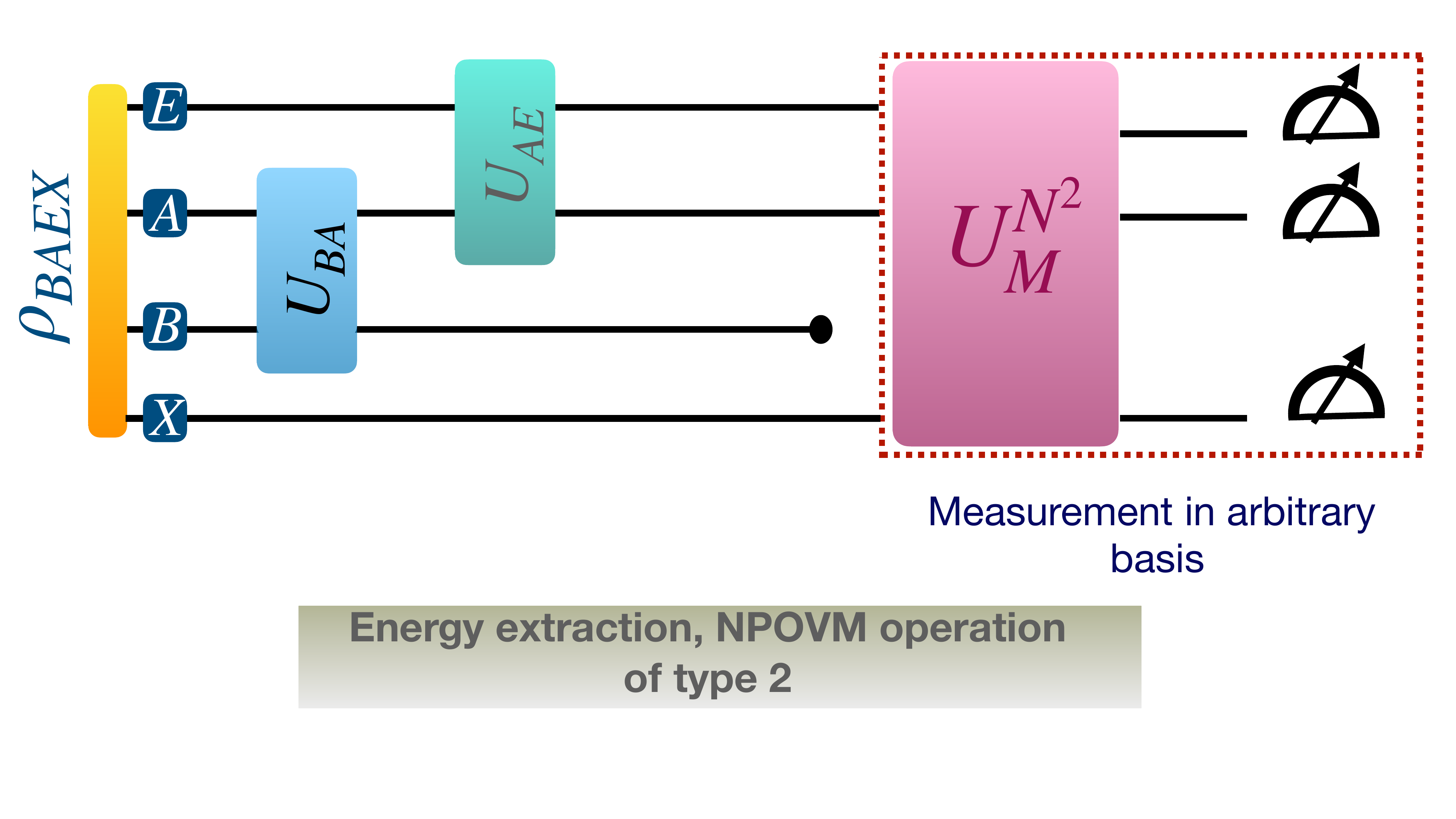 }
\caption{{\textbf{Description of the positive and non-positive operator-valued measurement-based energy extraction methods.} 
Two types of schematic diagrams are presented 
which represent our protocol. The upper diagram describes all operations in detail, while the lower one provides an analogous circuit representation.
    }}
		\label{figx3} 
		\end{figure*} 
  
The total change in the energy of the battery in the entire process is given by, $\Delta E = \text{tr}[\rho_{B} H_B]-\text{tr}[H_B \text{tr}_{AX}\left(\rho_{BAX}^{3,i}\right)]/p_i$, where $\text{tr}[\rho_B H_B]$ 
is the initial energy of the battery and $\rho_{BAX}^{3,i}$ is the final unnormalized state of $BAX$ after measurement, i.e., $\rho_{BAX}^{3,i}=\left(\mathbb{I}_B \otimes \ket{\Psi_i}_{AX}\bra{\Psi_i}\right)\text{tr}_E\big(\rho_S^2\big) \left(\mathbb{I}_B\otimes \ket{\Psi_i}_{AX}\bra{\Psi_i}\right) $ and $p_i=\text{tr}(\rho_{BAX}^{3,i})$ is the probability of getting the measurement outcome, $\ket{\Psi_i}_{AX}\bra{\Psi_i}$.

 The distillable energy, $S^P_{D}$, in this method, can be found by considering only the outcomes that provides positive extractable energy and optimizing over the set of all 
projectors, $\mathcal{M}$, on the Hilbert space describing jointly $AE$, i.e., 
\begin{eqnarray}
S^P_{D}&=&\max_{\ket{\Psi_i}_{AX}\in \mathcal{M}} \nonumber\sum_{i\in\mathcal{I}}\Big(p_i\text{tr}[\rho_{B} H_B] \\\nonumber
&& -
\;\; \text{tr}\left[H_B \text{tr}_{AX}\left(\rho_{BAX}^{3,i}\right)\right]\Big)\\\nonumber
&=&\max_{\ket{\Psi_i}_{AX}\in \mathcal{M}}\sum_{i\in\mathcal{I}}  \text{tr}[Z\rho_{{BAX}}^{3,i}]\nonumber,
\end{eqnarray}

 where $Z=\text{tr}[\rho_{B} H_B]\mathbb{I}_{BAX}-H_B \otimes \mathbb{I}_{AX}$. 
We can write
$\ket{\Psi_i}_{AX}=\widetilde{U}_{AX} \ket{0}$, where $\ket{0}$ is any fixed pure state of the system, $AX$, and can further consider the optimization involved in the expression of $S^P_{D}$ as a maximization over the set of all unitaries, $\{\widetilde{U}_{AX}\}$. By further simplification of the expression of $S^P_{D}$, we finally get 
\begin{eqnarray}\nonumber
S^P_{D}&=&\max_{\widetilde{U}_{AX}}\sum_{i\in\mathcal{I}}~\text{tr}[\widetilde{U}_{AX} \mathcal{A}_i \widetilde{U}_{AX}^{\dag}\mathcal{B}]\\\nonumber
&=&\text{tr}~[\max_{\widetilde{U}_{AX}}\widetilde{U}_{AX} \sum_{i\in\mathcal{I}}\mathcal{A}_i \widetilde{U}_{AX}^{\dag}\mathcal{B}],  
\end{eqnarray}
 where $\{\mathcal{A}_i\}$ is the set of measurement basis elements in computational basis that acts on $AX$ and $\ket{0}\bra{0}$ is one of the element of the mentioned set.
On the other hand the operator $\mathcal{B}$ denotes $\text{tr}_B\left[\left(\text{tr}_E\left(\rho^2_{S}\right)Z\right)\right] $.
The maximum would be reached for that $\widetilde{U}_{AX}=\widetilde{U}_{max}$ for which $[\widetilde{U}_{max} \sum_{i\in\mathcal{I}}\mathcal{A}_i \widetilde{U}_{max}^{\dag},\mathcal{B}]=0$, and if the set of eigenvalues $\{\beta_i\}$, of $\mathcal{B}$ satisfies $\beta_i\leq \beta_j$, then 
the eigenvalues $\{\alpha_i\}$, of $\widetilde{U}_{max} \sum_{i\in\mathcal{I}}\mathcal{A}_i\widetilde{U}_{max}^{\dag}$ should satisfy $\alpha_i\leq\alpha_j$. Thus, the expression of $S^P_{D}$ reduces to
$S^P_{D}=\sum_{i\in\mathcal{I}} \alpha_i \beta_i$, where $\{\alpha_i\}$ and $\{\beta_i\}$ follow the same order, as mentioned above. Since $\mathcal{A}_1, \mathcal{A}_2, \mathcal{A}_3, \mathcal{A}_4$ are rank-one projectors, it has only one non-zero eigenvalue and that is equal to unity which is clearly the largest among the set.
Moreover, we know the action of unitary operation on a state cannot change the eigenvalues of the state, and hence  $\widetilde{U}_{max} \sum_{i\in\mathcal{I}}\mathcal{A}_i \widetilde{U}_{max}^{\dag}$
will have the same eigenvalues as $\sum_{i\in\mathcal{I}}\mathcal{A}_i$. Now, as the eigenvalues of $\sum_{i\in\mathcal{I}}\mathcal{A}_i$ are $1$ or $0$, so only the positive eigenvalues${\{\beta_i\}}$ will be incorporated in the summation of $\sum_{i\in\mathcal{I}} \alpha_i \beta_i$. Thus, we have
\begin{equation}
S^P_{D}=\sum_{i,\beta_i>0}\beta_{i}, \label{eq1x}
\end{equation}
where the summation runs over those indices, $i$, for which $\beta_i>0$.
Therefore, the distillable energy $S^P_{D}$ is equal to the sum of all positive eigenvalues of $\mathcal{B}$. 
 
 {On the other hand, the maximum stochastically extractable energy where we consider energy extraction from a single outcome is given by
\begin{eqnarray}
S^P_{max}&=&\max_{\ket{\Psi_i}_{AX}\in \mathcal{M}} \left\{ \nonumber p_i\text{tr}[\rho_{B} H_B]-\text{tr}\left[H_B \text{tr}_{AX}\left(\rho_{BAX}^{3,i}\right)\right] \right\}\\\nonumber
&=&\max_{\ket{\Psi_i}_{AX}\in \mathcal{M}} \text{tr}[Z\rho_{{BAX}}^{3,i}]\nonumber,
\end{eqnarray}
By further simplification of the expression of $S^P_{max}$, we finally get 
\begin{eqnarray}\nonumber
S^P_{max}&=&\max_{\widetilde{U}_{AX}}\text{tr}[\widetilde{U}_{AX} \mathcal{A}_i \widetilde{U}_{AX}^{\dag}\mathcal{B}],
\end{eqnarray}
where $\mathcal{A}$ and $\mathcal{B}$ are defined previously.
The maximum would be reached for that $\widetilde{U}_{AX}=\widetilde{U}_{max}$ for which $[\widetilde{U}_{max} \mathcal{A}_i \widetilde{U}_{max}^{\dag},\mathcal{B}]=0$, and if the set of eigenvalues, $\{\beta_i\}$, of $\mathcal{B}$ satisfy $\beta_i\leq \beta_j$, 
then the eigenvalues, $\{\alpha_i\}$, of $\widetilde{U}_{max} \mathcal{A}_i\widetilde{U}_{max}^{\dag}$ should satisfy $\alpha_i\leq\alpha_j$. Thus the expression of $S^P_{max}$ reduces to
$S^P_{max}=\sum_i \alpha_i \beta_i$, where $\{\alpha_i\}$ and $\{\beta_i\}$ follow the same order, as mentioned above. Since $\mathcal{A}_1=\ket{00}\bra{00}$ is a pure state, it has only one non-zero eigenvalue and that is equal to unity which is clearly the largest among the set.
Moreover, we know the action of unitary operation on a state can not change the eigenvalues of the state, hence  $\widetilde{U}_{max} \mathcal{A}_1 \widetilde{U}_{max}^{\dag}$
will have the same eigenvalues as $\mathcal{A}_1$. Thus we have
\begin{equation}
    S^P_{max}=\beta_{max}, \label{eq1}
\end{equation}
where $\beta_{max}$ is the largest eigenvalue of $\mathcal{B}$. 
At the same time, the maximum average extractable energy by POVM operation is given by
\begin{eqnarray}
S^P_{av}&=&\max_{\ket{\Psi_i}_{AX}\in \mathcal{M}} \nonumber\sum_{i}\Big(p_i\text{tr}[\rho_{B} H_B] \\\nonumber
&& -
\;\; \text{tr}\left[H_B \text{tr}_{AX}\left(\rho_{BAX}^{3,i}\right)\right]\Big)\\\nonumber
&=&\max_{\ket{\Psi_i}_{AX}\in \mathcal{M}}\sum_{i}  \text{tr}[Z\rho_{{BAX}}^{3,i}]\nonumber,
\end{eqnarray}
{where the notations have their previous meanings.
Similarly, we can write the expression of $S^P_{av}$ in terms of the computational basis elements $\{\mathcal{A}_1, \mathcal{A}_3, \mathcal{A}_3,\mathcal{A}_4\}$, $\tilde{{U}}_{max}$ and $\mathcal{B}$, which is given by}
\begin{eqnarray}\nonumber\label{3x3}
S^P_{av}&=&\max_{\widetilde{U}_{AX}}\text{tr}[\widetilde{U}_{AX} \sum_i\mathcal{A}_i \widetilde{U}_{AX}^{\dag}\mathcal{B}]\\\nonumber
&=&\text{tr}[\widetilde{U}_{AX}  \widetilde{U}_{AX}^{\dag}\mathcal{B}]\\
&=&\text{tr}[\mathcal{B}]=\sum_i\beta_i.
\end{eqnarray}. The final expression in Eq.~\eqref{3x3}, comes from the completeness relation of the measurement operators, i.e.  $\sum_i\mathcal{A}_i=\mathbbm{I}$. {Therefore, the average extractable energy under a POVM operation no longer depends on the measurement setting; it simply equals to the difference between the battery’s initial energy and its energy just before the measurement is applied.}} 
 \end{proof}
\section{Implementing NPOVM on the auxiliary}\label{6S}
Here we consider two methods for implementation of NPOVM on the auxiliary. In the first case, the external remains merely a spectator, while in the second case, the external participates in the process.

\subsection{{Constructing} NPOVM excluding the external (type-1 NPOVM)}\label{6s1}

{In this subsection we provide the analytical form of distillable energy, maximum stochastically extractable energy, and average energy using NPOVM of type-1.}


{
For the application of the measurements, in this case, the initial states of $BAE$, $X$, and $S$  are considered to be $\rho^2_{BAE}$, $\rho_X$, and $\rho^2_S=\rho_{BAE} \otimes \rho_X$, respectively, just before the performance of the measurement. The battery's energy is defined by the Hamiltonian, $H_B$.}
{We can define the following operators: $\rho_B = \operatorname{tr}_{AE}[\rho_{BAE}^{\,2}]$,
$Z'=\text{tr}[\rho_{B} H_B]\mathbb{I}_{BAE}-H_B \otimes \mathbb{I}_{AE}$,
and
$\mathcal{B}' = \text{tr}_B\left[\rho^2_{BAE}Z'\right]$, where $\mathbb{I}$ denotes the identity operator acting on the relevant subsystem. The set of eigenvalues of $\mathcal{B}'$ can be denoted as $\{\beta_i'\}_i$.} 

\begin{theorem}
{Upon performing the optimal NPOVM of type-1 on $A$, the distillable, maximum stochastically extractable, and average extractable energies will be, respectively,
$S^{P}_D = \sum_{i \in \mathcal{I}} \beta'_i$, 
$S^{P}_{\max} =\max_i \beta'_{\max}$, 
and 
$S^{P}_{av} = \sum_i \beta'_i$,
where $\mathcal{I}$ denotes the set of indices denoting the positive eigenvalues among $\{\beta'_i\}$.}
\end{theorem}




\begin{proof}
To perform the NPOVM on $A$, instead of applying projective measurement on $AX$, we probabilistically project $AE$ on a measurement basis, $\{\ket{\Psi_i}_{AE}\}_i$. Since $AE$ is generally entangled, the projective measurement acting on $AE$ is equivalent to NPOVM applied on $A$. We refer to this as type-1 NPOVM.
 The amount of stochastically extracted energy from $B$ by performing a particular NPOVM of type-1 is given as, $S^{NP_1} =\sum_{i\in\mathcal{I}} p'_i\text{tr}[\rho_{B} H_B]-\text{tr}[H_B \text{tr}_{AE}(\rho_{BAE}'^{3,i})]$, where $\rho_{BAE}'^{3,i}=(\mathbb{I}_B \otimes \ket{\Psi_i}_{AE}\bra{\Psi_i})\rho_{BAE}^2(\mathbb{I}_B\otimes \ket{\Psi_i}_{AE}\bra{\Psi_i})$.
 
 {The distillable energy corresponding to NPOVM type-1 can also be written as \begin{eqnarray}
     S^{NP_1}_{D}
     &=&\max_{\widetilde{U}_{AE}} \sum_{i\in\mathcal{I}} p'_i\text{tr}[\rho_{B} H_B]-\text{tr}[H_B \text{tr}_{AE}(\rho_{BAE}'^{3,i})]\\
     &=&\max_{\widetilde{U}_{AE}} \sum_{i\in\mathcal{I}}\text{tr}\left(\widetilde{U}_{AE} \ket{0'}\bra{0'}\widetilde{U}_{AE}^{\dag}\text{tr}_B\left[\rho_{BAE}^2Z'\right]\right).\nonumber
 \end{eqnarray}} Here $Z'=\text{tr}[\rho_{B} H_B]\mathbb{I}_{BAE}-H_B \otimes \mathbb{I}_{AE}$ and $\ket{\Psi_i}_{AE}={\widetilde{U}_{AE}}\ket{0'}$ where $\ket{0'}$ is any fixed pure state of $AE$. Using the same logic as used in the case of POVM, the above expression can be further simplified to
{ \begin{equation}
     S^{NP_1}_{D}
    =\sum_{i\in\mathcal{I}}\max_{\widetilde{U}_{AE}}~ \text{tr}[\widetilde{U}_{AE} \mathcal{A}{'}_i \widetilde{U}_{AE}^{\dag}\mathcal{B}{'}]
    =\sum_{i}\beta'_i, \label{eqxx2}
 \end{equation}}
   where $\{\mathcal{A}{'}_i\}$ is the set of measurement basis elements in the computational basis that acts on $AE$ and $\ket{0'}\bra{0'}$ is one of the element of the set mentioned set. The expression of  $\mathcal{B}
{'}$ is given by $\text{tr}_B\left[\rho_{BAE}^2Z'\right]$, and $\beta'_{x}$ are the eigenvalues of the  operator$\mathcal{B}'$. Similar to the case of POVM operation we can also write the expression of maximum stochastically extractable energy and average energy in terms of the eigenvalues of the $\mathcal{B}$ operator is given by $S^{NP_1}_{max}=\beta'_{max}$ and $S^{NP_1}_{av}=\sum_{i}\beta'_i$.  $\beta'_{max}$ is the largest eigenvalue of $\mathcal{B}'$.
 \end{proof}


\subsection{{Constructing} NPOVM including the external (type-2 NPOVM)}\label{6s2}

{Similar to, NPOVM type-1 operation, we provide the analytical form of distillable energy, maximum stochastically extractable energy, and average energy using NPOVM of type-2.}


{Before the measurement is performed, the composite system $S=BAEX$ is described by the joint state $\rho_S$, which is assumed to be non-product with respect to the bipartition $BAE:X$. The battery is characterized by the Hamiltonian $H_B$.
We can define the following operators: $\rho_B = \operatorname{tr}_{AEX}[\rho_{BAEX}^{\,2}]$,
$Z''=\text{tr}[\rho_{B} H_B]\mathbb{I}_{BAEX}-H_B \otimes \mathbb{I}_{AEX}$,
and
$\mathcal{B}'' = \text{tr}_B\left[\rho^2_SZ''\right]$, where $\mathbb{I}$ denotes the identity operator acting on the relevant subsystem. The set of eigenvalues of $\mathcal{B}''$ can be denoted as $\{\beta_i''\}_i$.}

\begin{theorem}\label{th_3}
{The application of the optimal type-2 NPOVM on the subsystem $A$ yields the distillable, maximum stochastically extractable, and average extractable energies, respectively, as
$S^{P}_D = \sum_{i \in \mathcal{I}} \beta''_i$, 
$S^{P}_{\max} =\max_i \beta''_{\max}$, 
and 
$S^{P}_{av} = \sum_i \beta''_i$,
where $\mathcal{I}$ denotes the set of indices denoting the positive eigenvalues among $\{\beta''_i\}$.}
\end{theorem}

The proof of theorem~\ref{th_3}, is provided in appendix~\ref{AA_1}.

{
One can notice, while determining the distillable energy optimized over POVMs, we have maximized the quantity, keeping the dimension as well as the state of the external, arbitrary but fixed. Hence, the collection of POVMs considered here does not form the most general set. The same has been done for the case of distillable energy considering both types of NPOVMs. In particular, the distillable energy with NPOVMs is found keeping the environment state fixed and optimizing over the applied measurement basis. Moreover, to optimize over this measurement basis, we have introduced the set of all unitaries (i.e., $\{\widetilde{U}_{AE}\}$) of the considered dimension, which rotates the direction of the basis. As a result, the optimization involves measurements, which can effectively turn out to be POVM on auxiliary. However, for such a consideration, it can be shown that the taken set of POVMs to find $S_D^P$ is not the subset of the considered NPOVM set over which the optimization is performed to determine $S_D^{NP_1}$. The same remains true for the optimization involved to find maximum stochastically extractable energy. The proof is provided in appendix~\ref{aD}.
}





{\subsection{Maximum extractable energies using type-1 and type-2 NPOVM are noise-independent }\label{7s}}
{In this section we 
analyze an interesting feature of 
type-1 and type-2 
NPOVMs.
The two figures of merit that we 
consider, i.e. distillable energy and maximum probabilistically extractable energy, 
are independent of the type and strength of the noise applied.
\begin{theorem}\label{th_4c}
    The distillable energy ($S_D^{NP_{1/2}}$), the maximum stochastically extractable energy ($S_{max}^{NP_{1/2}}$), and the average energy ($S_{av}^{NP_{1/2}}$) using both type-1 and type-2 of NPOVMs are independent of the presence of noise.
\end{theorem}

The proof of theorem~\ref{th_4c} is provided in appendix~\ref{app_A}.

{The physical reason behind this noise independence is that when we perform POVM on the auxiliary system, we apply a global measurement on the joint auxiliary-external systems and then ignore the external system. In such a scenario, we do not have any control over the environment. As a result, we are unable to control the noise that affects the auxiliary because of its interaction with the environment. On the other hand, to perform the NPOVM (be it type-1 or type-2), we consider that the control over the environment is available. Therefore, in this case, we can choose a suitable direction of measurement that can completely nullify the effect of the noise.}

\section{extraction from single
qubit battery using POVM and type-1 NPOVM}\label{8S}

 \begin{figure}
		\centering
			\includegraphics[width=7.6cm]{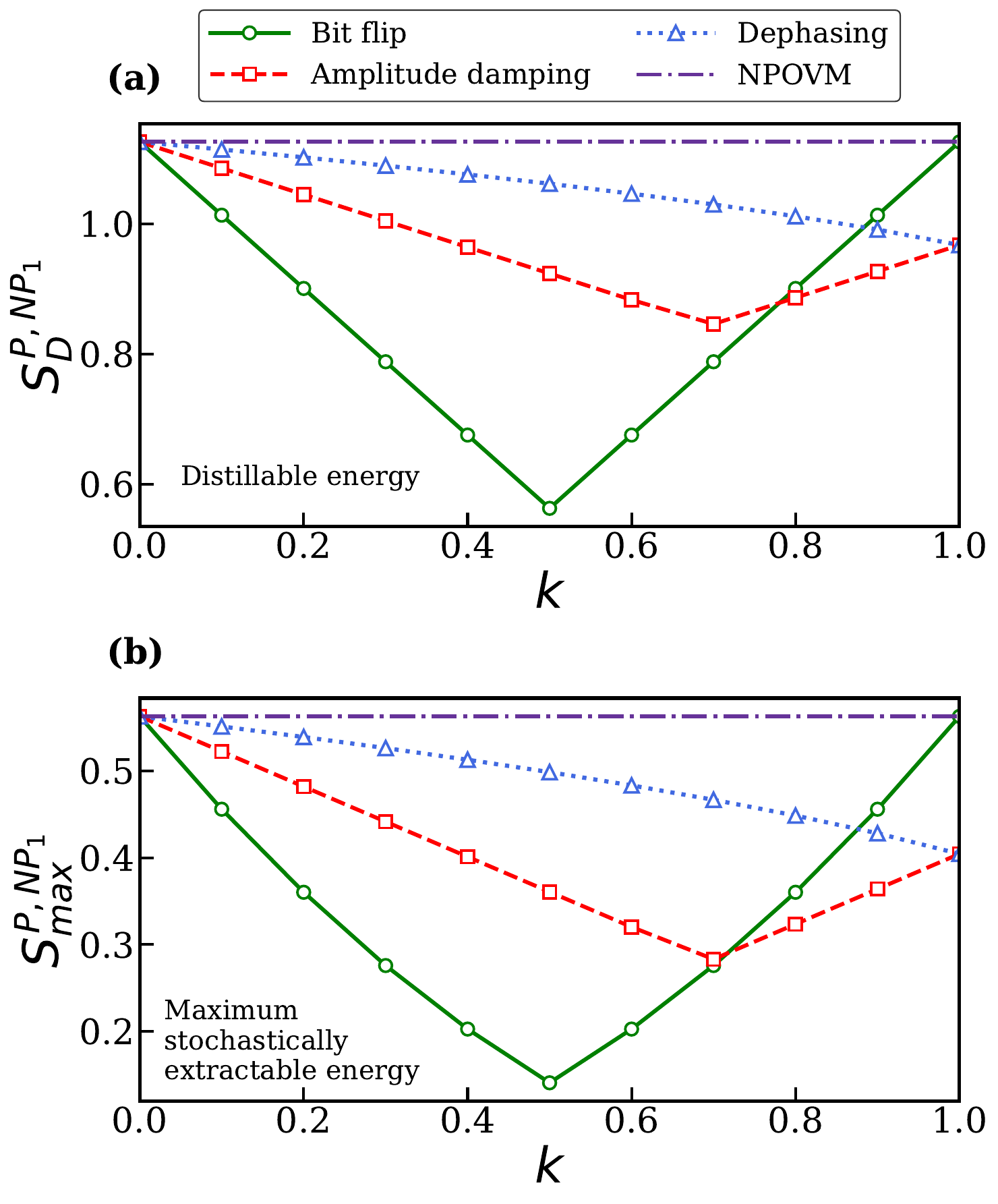}
\caption{
\textbf{Comparison between POVM and NPOVM of type-1 for three different noise models. } Panel (a) illustrates the behavior of  distillable energy extracted from a qubit battery by performing POVM and type-1 NPOVM measurements on an attached auxiliary qubit. We separately analyze the effects of the three different types of noise, namely amplitude damping, dephasing, and bit-flip, acting on the auxiliary qubit prior to the measurement. The quantities $S_{D}^{NP_1}$ and $S_{D}^{P}$ are plotted along the vertical axis as functions of the noise strength $k$, shown on the horizontal axis. {The smooth green curve with circular markers, the red dashed curve with square markers, and blue dotted  curve with triangular markers represent $S_{D}^{P}$ corresponding to bit-flip, amplitude damping, and dephasing noise, respectively, while the dashed dotted purple curve represents $S_{D}^{NP_1}$ for all three noise channels, as it has the same value for all noise models.}
The panel (b) presents the corresponding behavior of the stochastically extractable energy obtained from the qubit battery under POVM and type-1 NPOVM measurements performed on the auxiliary system. {Here, the line and marker styles and colors of the curves for the three different noise models and for both types of measurements are chosen to be the same as panel (a).} In both panels, the parameters are chosen as $J_{BA}=2h_A$, $h_B=h_A$, and $t=0.3h_A/J_{BA}$. The vertical axis is expressed in units of $h_A$, while the horizontal axis is dimensionless.}
		\label{fig33}
		\end{figure}

As the simplest non-trivial example, we choose the battery ($B$), auxiliary ($A$), environment ($E$), and external ($X$) to be qubits. The local Hamiltonian of $B$ ($A$/$E$) is taken as $H_{B(A/E)}=h_{B(A/E)}\sigma^z$, where $\sigma_z$ is a Pauli matrix. We consider both $B$ and $A$ to be initially prepared in the excited states, i.e., respectively, $\ket{e}_B$ and $\ket{e}_A$, whereas $E$ is taken to be in the ground state, $\ket{g}_E$. Hence the joint state of $BAE$ is given by $\ket{e}_B\bra{e} \otimes \ket{e}_A\bra{e}\otimes \ket{g}_E\bra{g}$.
At this point, we switch on an interaction for time $t$ between $B$ and $A$ described by the Hamiltonian $H_I=J_{BA}(\sigma^x \otimes \sigma^x)$, where $\sigma^x$ is the Pauli spin matrix and  $J_{BA}$ denotes the strength of the interaction. Hence the evolution  in between the time $(0,t]$ is governed by the Hamiltonian, $H_{BA}=H_B\otimes\mathbb{I}_A+\mathbb{I}_B\otimes H_A+H_I=h_B(\sigma^z\otimes \mathbb{I}_A)+h_A(\mathbb{I}_B\otimes \sigma^z)+J_{BA}(\sigma^x \otimes \sigma^x)$.
 In all the numerical calculations, we will, for specificity, consider $J_{BA}=2h_A$, and $h_B=h_A$. 
 
The evolution, governed by $H_{BA}$, will not affect $E$ which is initially prepared in the state $\ket{g}_E$. Therefore, the final state of $BAE$ after this interaction is $U_{BA}\left(\ket{e}_B\bra{e} \otimes \ket{e}_A\bra{e}\right)U_{BA}^\dagger\otimes \ket{g}_E\bra{g}$, where in this case $U_{BA}=\exp(-iH_{BA}t/\hbar)$. After time $t$, we evolve the joint state of $AE$ using a unitary $U_{AE}^i(k)$. Now we introduce the external to implement the POVM on $A$. To make an unbiased comparison between the performance of the POVM and  type-1, we consider the initial state of the external to be $\rho_X=\text{tr}_{BA}[\rho_{BAE}^2]$ (one needs to be careful with the notation here; on the right-hand side of the equation we present the state of $E$, and a copy of the same state is being prepared as $X$). $U_{AE}^i(k)$ is chosen in such a way that the corresponding evolution of $A$ can effectively be described by a amplitude damping (for $i=1$), bit flip (for $i=2$), or dephasing channel (for $i=3$), having strength $k$.

  The unitary representations, $U_{AE}^i(k)$, for the different channels are given by
 
  \begin{itemize}
      \item \textbf{{Amplitude damping channel:}}
{\begin{eqnarray}
&U_{AE}^1(k)\ket{e}_{A}\ket{g}_{E}&=\sqrt{1-k} \ket{e}_{A}\ket{g}_{E}+\sqrt{k}{\ket{g}_{A}\ket{e}_{E}},\nonumber\\ &U_{AE}^1(k)\ket{g}_{A}\ket{g}_{E}&=\ket{g}_{A}\ket{g}_{E},\nonumber 
\end{eqnarray}}

 \item \textbf{{Bit flip channel:}}
{\begin{eqnarray}
&U_{AE}^2(k)\ket{e}_{A}\ket{g}_{E}&=\sqrt{1-k} \ket{e}_{A}\ket{g}_{E}+\sqrt{k} \ket{g}_{A}\ket{e}_{E},\nonumber\\
&U_{AE}^2(k)\ket{g}_{A}\ket{g}_{E}&=\sqrt{1-k}\ket{g}_{A}\ket{e}_{E}+\sqrt{k}\ket{e}_{A}\ket{g}_{E},\nonumber
\end{eqnarray}}

 \item \textbf{{Dephasing channel:}}
 {\begin{eqnarray}
      &U_{AE}^3(k)\ket{e}_{A}\ket{g}_{E}&=\sqrt{1-k} \ket{e}_{A}\ket{g}_{E}+\sqrt{k} \ket{g}_{A}\ket{g}_{E},\nonumber\\
&U_{AE}^3(k)\ket{g}_{A}\ket{g}_{E}&=\ket{g}_{A}\ket{g}_{E}.\nonumber
\end{eqnarray}}
  \end{itemize}

Finally, we perform POVM and {NPOVM of type-1} on the auxiliary system and then calculate $S^{P}_{D}$ and $S^{NP_1}_{D}$ using Eqs. \eqref{eq1x} and \eqref{eqxx2}. We compare between $S^{P}_{D}$ and $S^{NP_1}_{D}$ by varying the noise parameter, $k$, through the illustration presented in the panel (a) of {Fig. \ref{fig33}}. {The smooth green
curve with circular markers, the red dashed curve with square
markers, and blue dotted curve with triangular markers denote $S_{D}^P$ for the cases where the applied noises are bit-flip, amplitude damping and dephasing, respectively.} As we have mentioned earlier, $S_{D}^{NP_1}$, does not depend on the applied noise, it is the same for all of the three considered noise models and is depicted using the {purple-colored dashed dotted lines}. {The flatness observed across all three types of noise 
at the same numerical value 
indicates that the distillable energy under NPOVM type-1 operations is independent of both the noise strength and the noise type.}
From these numerical studies, we conclude that distillable energy by NPOVM is more beneficial than using POVM. 
{The same numerical analysis is also performed in terms of the stochastically extractable energy, as shown in panel (b) of Fig.~\ref{fig33}.  The line and marker style and colors of curves for three different noise models and for both type of measurements are chosen same as panel (a).
The advantage of NPOVM type-1 operations over POVM is clearly evident in this case as well. Furthermore, our numerical results indicate that the stochastically extractable energy under NPOVM type-1 operations remains independent of both the type and the strength of the noise. }

 
{
  In Fig.~\ref{figx5}, we plot the variation of $S^{NP_1}_{D} - S^{P}_{D}$ as a function of $t$, considering the presence of amplitude damping noise implemented through $U^{1}_{AE}$ with noise strength $k = 0.5$. We observe that $S^{NP_1}_{D} - S^{P}_{D} \geq 0$ throughout.
  In other words, for the amplitude damping channel, the distillable energy obtained via the NPOVM operation is always greater than or equal to that obtained via the POVM operation, for all considered times $t$ at the fixed noise strength that is considered.
  The benefits in terms of stochastically extractable energy are also the same and shown in the inset of Fig.~\ref{figx5}, where $S^{NP_1}_{max}-S^{P}_{max}$ is plotted along the vertical axis with a navy blue colored curve having a square gray colored marker, and the horizontal axis represents time.

  {Though all the numerical results showing the advantage of NPOVMs are found considering qubit battery and auxiliary systems, to find them we have not considered any special property unique to two-level systems. We have restricted to qubit systems solely because of numerical simulation limitations. Therefore, intuitively we are optimistic that this advantage will also persist in higher-dimensional systems.}

}

 {It is to be noted that when we optimize over the POVMs and NPOVMs of both types, we only perform the optimization over the projective measurement in extended space but do not optimize over the initial states of the constituents involved in the protocol.}
 \begin{figure}
		\centering
			\includegraphics[width=7.6cm]{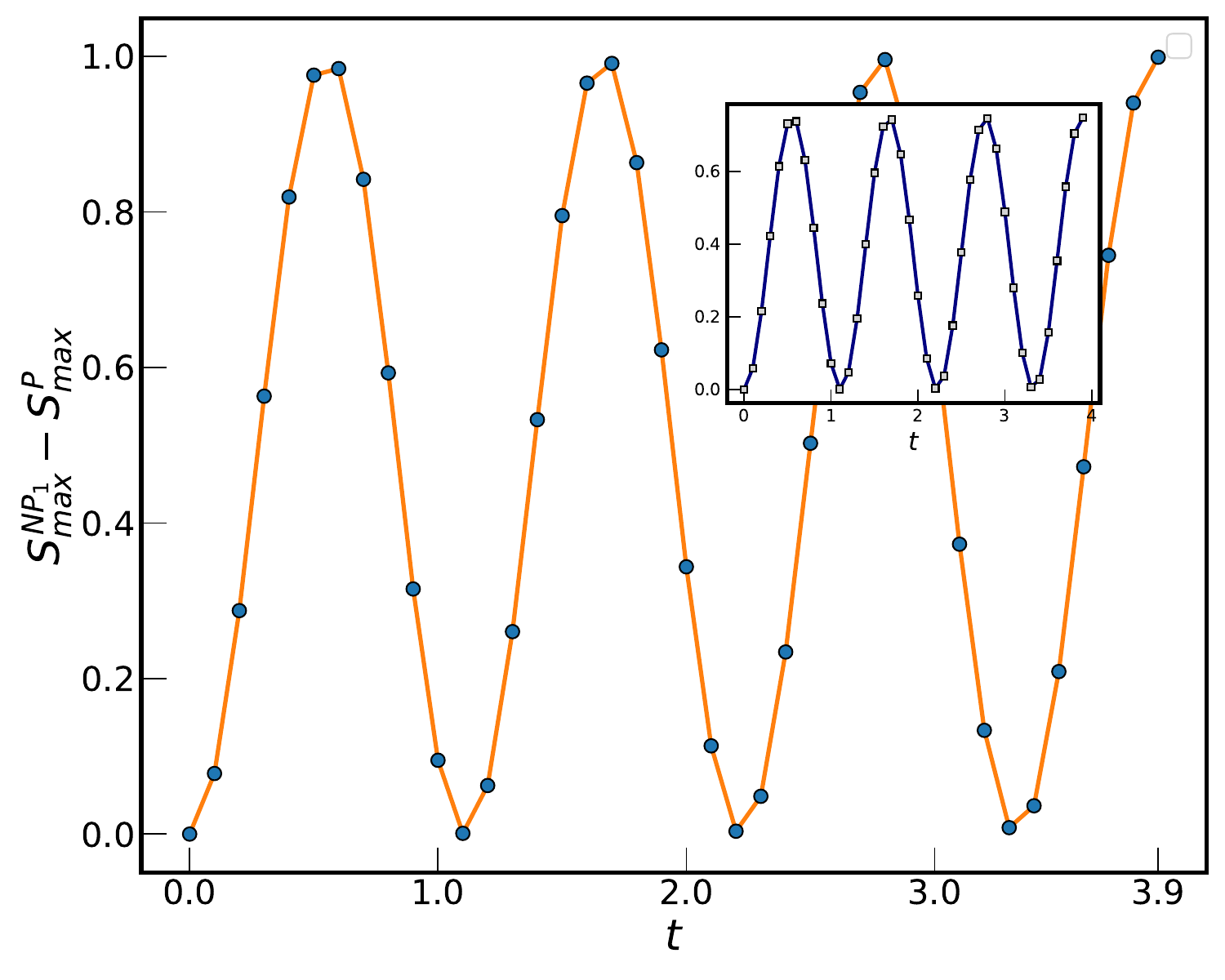}
\caption{\textbf{{Distillable and stochastically extractable energies of a battery by performing POVM and type-1 NPOVM on a connected auxiliary.}} { 
Along the vertical axis, we plot the difference between $S_{D}^{NP_1}$ and $S_{D}^{P}$ as a function of the interaction time $t$ between subsystems $B$ and $A$, which occurs in the first step of the measurement-based energy extraction process. This is shown by the orange curve with blue circular markers. The inset displays the variation of $S_{max}^{NP_1} - S_{max}^{P}$ with respect to $t$, represented by a navy curve with gray square markers. The horizontal axes denote the interaction time $t$ in units of $\hbar / h_A$, while the vertical axes are expressed in units of $h_A$. The parameters used for the plots are $J_{BA} = 2h_A$, $h_B = h_A$, and $k = 0.5$.
}}
		\label{figx5}
		\end{figure}

\section{Comparison between the performance of type-2
NPOVM and POVM in energy extraction}\label{11S}
Here we consider the initial states of $BAE$ and $BAEX$ for applying the POVM and NPOVMs, respectively, as 3- and 4-party GHZ class states.  The general $n$-party GHZ class state is given by
     $\ket{GHZ}^{n}_l=\sqrt{\frac{1+l}{2}}\ket{\phi}^{\otimes n}+\sqrt{\frac{1-l}{2}}\ket{\phi^\perp}^{\otimes n}$,
 where $l\in[-1,1]$ and $\{\ket{\phi},\ket{\phi^\perp}\}$ is any pair of orthogonal qubit states. Here we take $\ket{\phi}=\ket{e}$ and $\ket{\phi^\perp}=\ket{g}$. For application of both POVM and NPOVM, $BAE$ is first evolved using the unitaries $U_{BA}$ and $U_{AE}$, where the structure of the former is the same as before, and the form of the latter is taken such that it represents amplitude damping channel acting on $A$ {is given by}
{  \begin{equation}
U_{AE}^1{(k)}=\begin{bmatrix}\label{2}
			1 & 0 & 0& 0\\
			0& \sqrt{1-k} & \sqrt{k}& 0 \\
   0& -\sqrt{k}& \sqrt{1-k} &0 \\
    0& 0& 0 &1 \\
		\end{bmatrix}.\nonumber
\end{equation}}

 To perform the POVM, we introduce the external, $X$, prepared in the state $\rho_X=\frac{1+l}{2}\ket{0}\bra{0}+\frac{1-l}{2}\ket{1}\bra{1}$, which is basically the single-qubit reduced density matrix of $\ket{GHZ}^n_l$. Finally, the POVM and type-2 NPOVMs are applied on $A$ by performing projective measurements on the subsystems $AX$ and $AEX$, respectively, of the entire system $S$. 
{We calculate {$S^P_{D}$} and {$S^{NP_2}_{D}$} using, Eqs. \eqref{eq1x} and \eqref{eq3x}, for amplitude damping noise, and depict them in panel (a) of Fig.~\ref{fig55}, along the vertical axis, against the noise strength, $k$, where $k\in\{0,1\}$, and the initial state parameter, $l$, that are presented along the horizontal axes.
One can notice from the figure, that except the values of  $\{\tilde{k},1\}$ $\forall$  $\tilde{k}\in[0,1]$ and $\{\tilde{k},-1\}$ $\forall$  $\tilde{k}\in[0,1]$,}
{the distillable energy by type-2 NPOVM operation is greater than that using the POVM. {
\label{sec4}
\begin{figure}
		\centering
			\includegraphics[width=8.4cm]{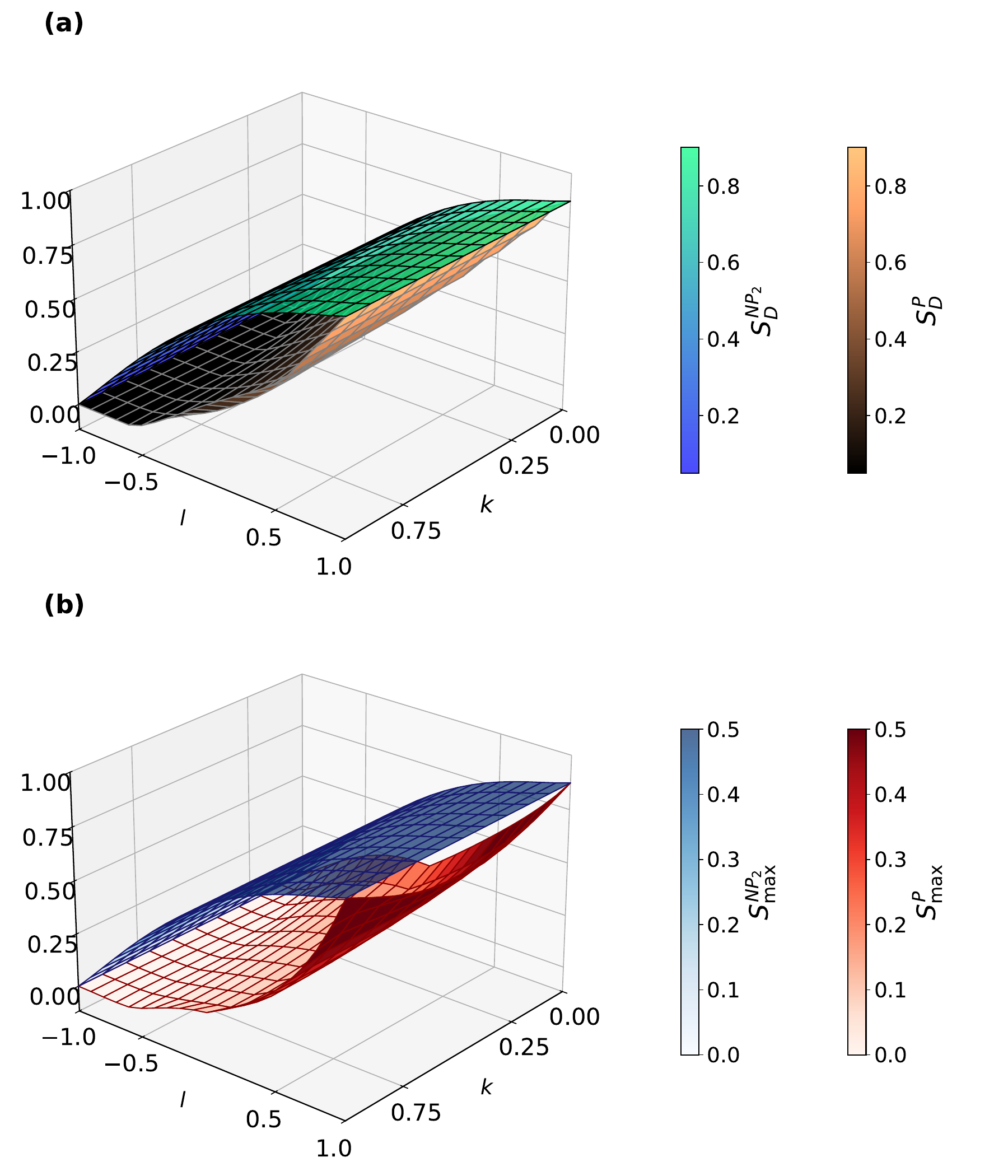}
\caption{{\textbf{Comparison between POVM and NPOVM of type-2 for amplitude damping noise.} Comparison between the distillable energies obtained using POVM and type-2 NPOVM is shown in panel (a). We plot $S_{D}^{NP_2}$ (bluish green surface) and $S_{D}^{P}$ (orange surface) along the vertical axis as functions of the noise strength $k$ and the parameter $l$ defining the initial state, displayed along the horizontal axes. The corresponding surface values are indicated by the color bars. From the figure, it is evident that the distillable energy achieved via type-2 NPOVM is consistently greater than that obtained using POVM for all values of $k$ and $l$. The vertical axis is expressed in units of $h_A$, while the horizontal axes are dimensionless.
The panel (b) presents the same comparison for the stochastically extractable energy. In this case, we plot $S_{\max}^{NP_2}$ (blue surface) and $S_{\max}^{P}$ (orange surface) as functions of $k$ and $l$. The results clearly demonstrate that the maximum stochastically extractable energy obtained through type-2 NPOVM is always greater than that achieved using POVM over the entire parameter range. As in the upper panel, the vertical axis is measured in units of $h_A$, and the horizontal axes are dimensionless. In both panels, the parameters are chosen as $J_{BA}=2h_A$, $h_B=h_A$, and $t=0.3h_A/J_{BA}$.}}
		\label{fig55}
		\end{figure} 
Similarly, we have carried out an analogous analysis for the stochastically extractable energy obtained via POVM and type-2 NPOVM operations, denoted by $S^{P}_{max}$ and $S^{NP_2}_{max}$, respectively, as shown in the panel (b) of Fig.~\ref{fig55}. It is also evident from the figure that, except at the points 
$\{0,1\}$, $\{1,1\}$, and along the line $\{\tilde{k},-1\}$ for all 
$\tilde{k} \in [0,1]$, the maximum stochastically extractable energy 
obtained via type-2 NPOVM operations exceeds that achievable using 
standard POVM operations.}}

\begin{figure}
		\centering
			\includegraphics[width=8.5cm]{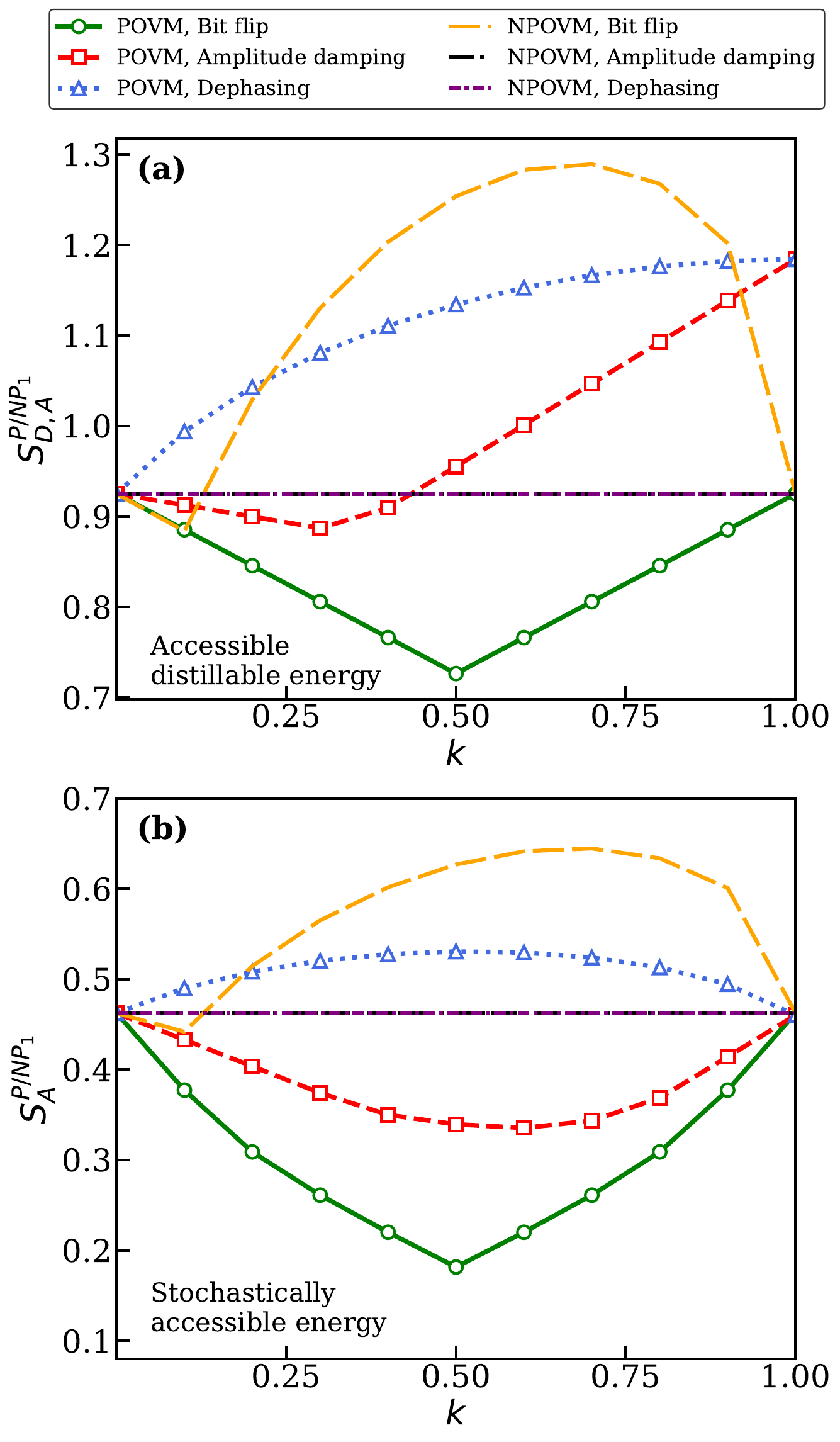}
\caption{{\textbf{Accessible distillable energy and stochastically accessible energy for various types of noise models.} In {panel (a)},
the behavior of accessible distillable energy from a qubit battery under POVM and type-1 NPOVM applied to an attached auxiliary system is shown in the main figure. Along the vertical axis, we plot $S_ {D,A}^{NP_1}$ and $S_{D,A}^P$ under bit-flip, amplitude-damping, and dephasing noise, as a function of the noise strength $k$, which is displayed on the horizontal axis. The {panel (b)} illustrates the accessible extractable energy from a qubit battery under POVM and type-1 NPOVM. Here, the vertical axis represents $S_{A}^{NP_1}$ and $S_{A}^P$ under bit-flip, amplitude-damping, and dephasing noise, again plotted against the noise strength $k$ shown along the horizontal axis. The remaining parameters are chosen as $J_{BA}=2h_{A}$, $h_{B}=h_{A}$, and $t=0.3h_{A}/J_{BA}$. The vertical axis is given in units of $h_{A}$, while the horizontal axis is dimensionless. The line styles corresponding to each curve {for both the panels} are specified in the legend.}}
		\label{xxx}
		\end{figure}

\section{Accessible distillable energy and Stochastically accessible energy
using POVM and NPOVM}\label{12S}

{In practical scenarios, preparing and performing the optimal measurement on the auxiliary qubit, $A$, i.e., the best projective measurement on $AE$ or $AX$ may be expensive. {The experimentalists, in such scenarios, may not have access to the optimal set of measurements.} To examine such situations, we perform the optimization involved in the definitions of {distillable energy corresponding to POVM and NPOVM type-1 operations, maximized over a limited set of measurement operators.} We refer to the resulting quantity as accessible energy, emphasizing that the NPOVM operations involved here are constructed only from the accessible measurement operators, rather than from the most general class of NPOVMs considered in the previous sections. In particular, in this restricted setting the nonlocal unitaries used to implement the NPOVM are assumed to arise solely from the interaction Hamiltonian between the system and the environment, which is taken to be the only nonlocal unitary available in the experimental setup. Consequently, the class of measurements considered here corresponds to those that can be physically realized using the natural system-environment interaction, thereby capturing experimentally accessible energy extraction protocols. The projective measurements applied on $AX$ (in case of POVM) and $AE$ (in case of NPOVM of type-1) can be implemented by operating unitaries on, respectively, $AX$ and $AE$, and then measuring the systems, $AX$ and $AE$, in the computational basis. Since we are already using a unitary, $U_{BA}=\exp(-iH_{BA}t/\hbar)$, to entangle $B$ and $A$ in the first step, we can assume that such unitary, $U_{BA}$, is available in the laboratory. Therefore we will use the same unitary, $U_{AE(AX)}=\exp(-iH_{AE(AX)}{t}/\hbar)$, on $AE$ ($AX$) to perform measurement on $AE$ ($AX$) and finally apply a local projective measurement, $\{\ket{\Psi_i}_A\ket{\Psi_j}_{E(X)}\}$ on $AE$ ($AX$). Here $H_{AE(AX)}$ is given by $H_{AE(AX)}=h_A(\sigma^z\otimes \mathbb{I}_{{E(X)}})+h_{E(X)}(\mathbb{I}_A\otimes \sigma^z)+J_{AE(AX)}(\sigma^x \otimes \sigma^x)$, which is the same Hamiltonian as $H_{BA}$ with the only difference that it is defined to act on the system, $AE$ ($AX$). }Since the form of the unitary, $U_{AE(AX)}$, is the same as $U_{BA}$, we expect that if $U_{BA}$ can be produced in the laboratory then $U_{AE(AX)}$ can also be constructed. Furthermore, since performing measurement in a product
basis, $\{\ket{\Psi_i}_A\ket{\Psi_j}_{E(X)}\}$, does not require any additional non-local resources we believe measurements in the basis, $\{\ket{\Psi_i}_A\ket{\Psi_j}_{E(X)}\}$, can also be easily implemented. 
Therefore, instead of optimizing $S^{NP_1}_d$ ($S^{P}_d$) over all unitaries, $\widetilde{U}_{AE(AX)}$, we optimize $S^{NP_1}_d$ ($S^{P}_d$) over a restricted set of unitaries, $\{\widetilde{U}_A\otimes \widetilde{U}_{E(X)}\}$, production of which involves lesser cost. 
The distillable energy and stochastically extracted energy optimized over this smaller set of unitaries is named accessible distillable energy and stochastically accessible energy, in the sense, that it is the amount of energy of the battery that can be accessed in the laboratory. {We denote the accessible distillable energy and the stochastically accessible energy by $S_{D,A}^{P}$ and $S_{D,A}^{NP_1}$, and $S_A^{P}$ and $S_A^{NP_1}$, corresponding to the cases where the auxiliary is subjected to a POVM and a type-1 NPOVM operation, respectively. In both cases, distillable and stochastically extractable energy, the superscripts $P$ and $NP_1$ denote POVM and type-1 NPOVM, respectively. Formally, the expressions of $S_{D,A}^{P}$ and $S_{D,A}^{NP_1}$ are given, respectively, by}

{\begin{eqnarray}\nonumber
S_{D,A}^{P}=\max_{\widetilde{U}_A,\widetilde{U}_{X} }\sum_{\ket{\psi_i}\in \mathcal{I}^{P}}\tr\Big(\widetilde{U}_{A}\otimes\widetilde{U}_{X} \ket{\psi}^i_{AX}\bra{\psi}^i_{AX}\widetilde{U}^\dagger_{A}\otimes\widetilde{U}^\dagger_{X}\\\nonumber
\tr_B\left[\left(\tr_E\left(\rho^2_{S}\right)Z\right)\right]) 
\end{eqnarray}}
and 
{\begin{eqnarray}\nonumber
S_{D,A}^{NP_1}=\max_{\widetilde{U}_A,\widetilde{U}_{E} }\sum_{\ket{\xi_i}\in \mathcal{I}^{NP_1}}\tr\Big(\widetilde{U}_{A}\otimes\widetilde{U}_{E} \ket{\xi}_{AE}\bra{\xi}_{AE}\widetilde{U}^\dagger_{A}\otimes\widetilde{U}^\dagger_{E}\\\nonumber
\tr_B\left[\rho^2_{BAE}Z'\right]\Big),
\end{eqnarray}}
 where  {$\ket{\psi}_{AX}=\exp(-iH_{AX}t/\hbar)\ket{0}_A\ket{0}_X$ and $\ket{\xi}_{AE}=\exp(-iH_{AE}t/\hbar)\ket{0}_A\ket{0}_E$.} Here $\ket{0}_A$, $\ket{0}_X$, $\ket{0}_E$, are any fixed pure states of $A$, $X$, and $E$, respectively. 
{
Let $\mathcal{I}^{NP_1}$ and $\mathcal{I}^{P}$ 
denote the sets of outcomes corresponding to POVM and NPOVM operations, respectively, for which the 
expressions,
$\tr\Big(\widetilde{U}_{A}\otimes\widetilde{U}_{X} \ket{\psi}^i_{AX}\bra{\psi}^i_{AX}\widetilde{U}^\dagger_{A}\otimes\widetilde{U}^\dagger_{X}
\tr_B\left[\left(\tr_E\left(\rho^2_{S}\right)Z\right)\right])$ and $\tr\Big(\widetilde{U}_{A}\otimes\widetilde{U}_{E} \ket{\xi}_{AE}\bra{\xi}_{AE}\widetilde{U}^\dagger_{A}\otimes\widetilde{U}^\dagger_{E}
\tr_B\left[\rho^2_{BAE}Z'\right]\Big)$ are positive.
The corresponding quantities, $S^{NP_1}_{A}$ and $S^{P}_{A}$, are given by}
{\begin{eqnarray}\nonumber
S_A^{P}=\max_{\widetilde{U}_A,\widetilde{U}_{X} }~\tr\Big(\widetilde{U}_{A}\otimes\widetilde{U}_{X} \ket{\psi}^i_{AX}\bra{\psi}^i_{AX}\widetilde{U}^\dagger_{A}\otimes\widetilde{U}^\dagger_{X}\\\nonumber
\tr_B\left[\left(\tr_E\left(\rho^2_{S}\right)Z\right)\right]) 
\end{eqnarray}
and 
\begin{eqnarray}\nonumber
S_A^{NP_1}=\max_{\widetilde{U}_A,\widetilde{U}_{E} }~\tr\Big(\widetilde{U}_{A}\otimes\widetilde{U}_{E} \ket{\xi}_{AE}\bra{\xi}_{AE}\widetilde{U}^\dagger_{A}\otimes\widetilde{U}^\dagger_{E}\\\nonumber
\tr_B\left[\rho^2_{BAE}Z'\right]\Big).
\end{eqnarray}}

{{The behaviors of $S^{NP_1}_{D,A}$ and $S_{D,A}^{P}$ are presented in panel (a) of Fig.~\ref{xxx} for each of the three different noises, i.e., bit-flip, amplitude damping, and dephasing noise, denoted by smooth green line with circular markers,  red dashed line with  square markers, and dotted blue line with triangular markers for the POVM operation and orange, black, and purple colored lines with different line style (mentioned in the legend) for the NPOVM type-1 operation for  bit-flip, amplitude damping, and dephasing noise.}  Here, the accessible distillable energies using NPOVM, in presence of amplitude damping and dephasing noise, coincides with one another.}

{
From the figure, it is evident that under dephasing noise, {the POVM is advantageous} in extracting accessible distillable energy for all values of the noise parameter $k$ except $k=0$ and $k=1$. Since the implementation of a POVM requires fewer resources, this advantage is achieved in a more resource-efficient manner.
In contrast, for bit-flip noise, type-1 NPOVM offers an advantage over POVM within a broad range of $k \in [0.1,1]$, while in the interval $k \in [0,0.1]$, both POVM and type-1 NPOVM yield the same amount of accessible distillable energy. Finally, in the case of amplitude-damping noise, type-1 NPOVM surpasses POVM in the region $k \in [0,4.29]$, whereas beyond this range, the resource-efficient POVM performs better than the type-1 NPOVM.}

{The stochastically accessible energy achieved by both POVM and NPOVM type-1 operations for bit-flip, amplitude damping, and dephasing noise is plotted along the vertical axis of the panel (b) of Fig.~\ref{xxx}, where the horizontal axis denotes the noise parameter $k$.
Similar to the accessible distillable energy, stochastically accessible energy by resource-efficient positive measurements outperforms that of NPOVM type-1 operation for all values of noise parameter except $k=0$ and $k=1$ for dephasing noise. {Here, the curves corresponding to different types of noise models have the same line and marker style and color as that of panel (a) for both POVM and NPOVM type-1 measurements.}
The stochastically accessible energy extracted by NPOVM can be seen to be independent of the noise strength for dephasing and amplitude damping noises.
On the other hand, in the case of bit flip and amplitude damping noise, there exists an advantage of using NPOVM over POVM. } \\

 \section{conclusion}\label{CCCCCCCCCCC}

Measurement-based method of energy extraction from quantum batteries was introduced in Ref.~\cite{demon1,ord_demon,auxi}, which involved action of POVM on an auxiliary attached to the battery. 
Here we identified the most general quantum mechanically allowed measurements, which could be NPOVMs and compared the effectiveness of NPOVM and POVM in measurement-based energy extraction from quantum batteries where the measurement under consideration is always performed on the attached auxiliary.

We first introduced two types of NPOVMs and derived the expressions for the {distillable energy,} maximum stochastically extractable energies, {and average extractable energy} using POVMs and NPOVMs of both types. Our method involves first switching on an interaction between the battery and the auxiliary, then applying a noise on the auxiliary by introducing an interaction between the auxiliary with its environment and then finally applying projective measurements on the joint system: auxiliary-external, auxiliary-environment, and auxiliary-environment-external, which represents the application of, respectively, POVM, type-1 NPOVM, and type-2 NPOVM, on the auxiliary. We proved that {distillable energy} and stochastically extractable energy using NPOVMs of both types does not depend on the applied noise on the auxiliary. {The average extractable energy is even independent of the applied measurement and is equal to the energy difference between the battery's initial state and the state just before the application of the measurement.} 

Subsequently, we focused on qubit batteries described by a particular Hamiltonian and compared the {distillable energy} and stochastically extractable energy using POVMs and {NPOVMs of type-1}. By separately considering amplitude damping, bit-flip, and dephasing noise, acting on the auxiliary we proved that the {distillable energy} and stochastically extractable energy using NPOVM of type-1 is always greater than or equal to the same for POVMs, whatever the strength of the considered noise. Moreover, considering a class of initial states we made a comparison in terms of {distillable energy} and maximum stochastically extractable energy between the using POVM and type-2 NPOVM in the presence of amplitude damping noise affecting the auxiliary. We show that also in this case, for both the {figure of merits} using NPOVM of type-2 is greater than or equal to that using POVM.

At last we dealt with situations where access to only a restricted set of POVM and type-1 NPOVM is available.
{In such a scenario we define two quantities, i.e., accessible distillable energy and stochastically accessible energy.}
By fixing the time interval of the interaction between the auxiliary and the battery, we found that in most of the situations NPOVM offers benefits over POVM. {However, we also observed instances in which resource-efficient POVMs are advantageous.} 

We have shown the presestance of advantage of NPOVMs in qubit  dimensional auxiliary and battery, the higher dimensional annalysis of battery and auxiliary system as a direction for future research.

{Having established the advantage of NPOVMs for qubit-dimensional battery and auxiliary systems, we identify the extension of our analysis to higher-dimensional battery and auxiliary systems as a promising direction for future research.}

\acknowledgements
{The research of PC was supported by the INFOSYS scholarship.}
KS acknowledges support from the project MadQ-CM (Madrid Quantum de la Comunidad de Madrid) funded by the European Union (NextGenerationEU, PRTRC17.I1) and by the Comunidad de Madrid (Programa de Acciones Complementarias).  {US acknowledges financial support from the Anusandhan National Research Foundation (ANRF), Government of India, under the Grant No. ANRF/ARG/2025/004617/PS.}

\onecolumngrid
\section*{Appendix}

\appendix

\section{Analytical form of distillable energy stochastically extractable energy, and average energy.}\label{AA_1}
Here we consider the initial state of the system to be $\rho^0_S=\rho_{BAEX}^0$. This state, $\rho^{0}_S$, evolves to  $\rho_S^1$ and then to the state, $\rho_S^2$, in the same way as before, i.e., $\rho_S^2=(\mathbb{I}_{B}\otimes {U}_{AE}\otimes \mathbb{I}_{X})(U_{BA}\otimes \mathbb{I}_{E}\otimes \mathbb{I}_{X}) \rho^{0}_S(U_{BA}^\dagger\otimes \mathbb{I}_{E}\otimes \mathbb{I}_{X})(\mathbb{I}_{B}\otimes {U}_{AE}^\dagger\otimes \mathbb{I}_{X})$, where one may need to keep in mind that here the form of $\rho_S^0$ is different from the one taken in the previous two sections as, in this case, $\rho_S^0$ may not be separable in the bipartition $BAE:X$. After obtaining the state $\rho_S^2$, an NPOVM is implemented on $A$
by performing projective measurement, $\{\ket{\Psi_i}_{AEX}\bra{\Psi_i}\}$, on the joint state of $AEX$. We name it as the type-2 NPOVM. The distilled energy from a quantum battery using a particular type-2 NPOVM is $S^{NP_2} = \sum_{i\in\mathcal{I}}p''_i\text{tr}[\rho_{B} H_B]-\text{tr}[H_{B _{AEX}}(\rho_{BAEX}''^{3,i})]$, where $\rho_{BAEX}''^{3,i}=(\mathbb{I}_B \otimes \ket{\Psi_i}_{AEX}\bra{\Psi_i})\rho_S^2(\mathbb{I}_B\otimes \ket{\Psi_i}_{AEX}\bra{\Psi_i})$ and $p''_i=\text{tr}(\rho_{BAEX}''^{3,i})$.
Our aim is to find out the stochastically extractable energy, $S^{NP_2}_{D}$, in this process, by maximizing over the set of all projectors, $\mathcal{M}''$, acting on the joint Hilbert space of $AEX$. The expression of $S^{NP_2}_{D}$ is given by
   $ S^{NP_2}_{D}= \max_{\ket{\Psi_i}_{AEX}\in \mathcal{M}''} \sum_{i\in\mathcal{I}}p''_i\text{tr}[\rho_{B} H_B]-\text{tr}[H_{B_{AEX}}(\rho_{BAEX}''^{3,i})]. $
By further simplifying the quantity, $S^{NP_2}_{D}$, we obtain
$S^{NP_2}_{D}=\max_{\widetilde{U}_{AEX}}\sum_{i\in\mathcal{I}}\tr\left(\widetilde{U}_{AEX} \ket{0''}\bra{0''}\widetilde{U}_{{AEX}^{\dag}} \tr_B\left[\left(\rho^2_{S}\right)Z''\right]\right)$.
Here $Z''=\text{tr}[\rho_{B} H_B] \mathbb{I}_{BAEX}-H_B \otimes \mathbb{I}_{AEX}$ and $\ket{\Psi_i}_{AEX}={\widetilde{U}_{AEX}}\ket{0''}$, where $\ket{0''}$ is any fixed pure state of $AEX$. Following the same calculations as in the case of POVM-based energy extraction, the stochastically extractable energy using type-2 NPOVM is found to be 
{
\begin{equation}
    S^{NP_2}_{D}
    =\max_{\widetilde{U}_{AEX}}~ \sum_{i\in\mathcal{I}}\text{tr}[\widetilde{U}_{AEX} \mathcal{A}{''}_i \widetilde{U}_{AEX}^{\dag}\mathcal{B}{''}]
    =\sum_{i\in\mathcal{I}}\beta''_{i}, \label{eq3x}
\end{equation}}
 where $\mathcal{A}{''}_i$ are three qubit measurement operator in computational basis that acts on the system $AEX$, where $\ket{0''}\bra{0''}$ is one of it's element  and  $\mathcal{B}
{''}$  is $\text{tr}_B\left[\left(\rho^2_{S}\right)Z''\right]$. $\{\beta_x\}$ is the the set of eigenvalues of the operator $\mathcal{B}$. 

Similar,here also, we can also write the expression of maximum stochastically extractable energy and average energy in terms of the eigenvalues of the $\mathcal{B}$ operator is given by $S^{NP_2}_{max}=\beta''_{max}$ and $S^{NP_2}_{av}=\sum_{i}\beta''_i$. $\beta''_{max}$ is the maximum eigenvalue of $\mathcal{B}''$.

\section{Maximum extractable energies using type-1 and type-2 NPOVM are noise-independent}\label{app_A}

In this section we 
provide the proof that distillable energy and maximum probabilistically extractable energy and average energy 
are independent of the type and strength of the noise applied when NPOVM type-1 and type-2 measures are performed.
In case of type-1 NPOVM operation, let us consider the initial state of composite system of battery (B), auxiliary (A), environment (E) and external system (X) to be $\rho^0_S=\rho^0_{BAE}\otimes\rho_X$. After the interaction between B and A through the unitary $U_{BA}$ the final state of $S$ is given by $\rho_{S}^1=(U_{BA}\otimes \mathbb{I}_{E}\otimes \mathbb{I}_{X}) \rho^{0}_S(U_{BA}^\dagger\otimes \mathbb{I}_{E}\otimes \mathbb{I}_{X})=\rho_{BAE}^1\otimes \rho_{X}$. Then the action of the unitary $U_{AE}$ on the auxiliary and environment, the final state is 
$\rho_S^2=\rho_{BAE}^2\otimes \rho_X$. Here, $\rho_{BAE}^2=(\mathbb{I}_B\otimes U_{AE})\rho_{BAE}^1(\mathbb{I}_B\otimes U_{AE}^\dagger)$, and 
the external is still fixed at $\rho_X$ and in a product state with rest of the system. Now the distillable energy after performing the NOPVM type-1 measurement on the auxiliary system realized through the applying global measurement on the auxiliary and environment and then tracing out the environment part is given by 

       \begin{eqnarray}\label{eq7}
         S^{NP_1}_{D}
&=&\max_{U_M^{N^2}}~\sum_{i\in\mathcal{I}} p'_i\text{tr}[Z'(\mathbbm{I}_B\otimes \left(U_M^{N^2}\right) \mathcal{A}'_i\left(U_M^{N^2}\right)^{\dag})(\mathbbm{I}_B\otimes U_{AE})
     (\rho_{BA}\otimes\rho_{E})(\mathbbm{I}_B\otimes \left(U_M^{N^2}\right)\mathcal{A}'_i\left(U_M^{N^2}\right)^{\dag})(\mathbbm{I}_B\otimes U^{\dag}_{AE})] \;\;\;\; \\\nonumber
     &=&\max_{U_M^{N^2}}~ \sum_{i\in\mathcal{I}} p'_i\text{tr}[Z'(\mathbbm{I}_B\otimes \left(U_M^{N^2}\right)U_{AE}^{\dag}\mathcal{A}'_i\mathcal{A}'^{\dag}_i\left(U_M^{N^2}\right)^{\dag}U_{AE})
     (\rho_{BA}\otimes\rho_{E})]. \nonumber
      \end{eqnarray}  

Here, the set $\{\left(U_M^{N^2}\right)\mathcal{A}'_i\left(U_M^{N^2}\right)^{\dag}\}$ represents the measurement operators acting jointly on the subsystems $A$ and $E$. Here, the optimization in Eq.~\eqref{eq7} is performed over all possible two-qubit unitaries $U_M^{N^2}$. This optimization can equivalently be expressed in terms of the composite operator $\widetilde{U}_{AE} = U_M^{N^2} U_{AE}$. 
We incorporate the unitary $U_{AE}$ into $\left(U_M^{N^2}\right)$ and define the composite unitary $\widetilde{U}_{AE}$, where $U_{AE}$ denotes the noisy operation acting on the auxiliary system and the environment. Consequently, $\widetilde{U}_{AE}$ spans the entire two-qubit unitary group, implying that $U_{AE}$ does not impose any restriction on the optimization. It implies that whatever be the noisy unitary $U_{AE}$ is, it has no effect on the optimization. Therefore, the maximum extractable energy under an NPOVM type-1 operation becomes independent of both the nature and the strength of the noise.
In a similar way, 
the maximum  stochastically extractable energy and average extractable energy also become independent of noise strength and the type of noise. 
Analogously, for NPOVM type-2 operation, we consider the initial state of the composite system of battery (B), auxiliary (A), environment (E) and external system (X) 
to be given by a general density operator $\rho_{BAEX}$, where the external system is not 
in a product state with rest of the system initially. After the action of $U_{BA}$ on the battery (B) and auxiliary system (A) the final state will be given by $\rho^1_{BAEX}=(U_{BA}\otimes\mathbbm{I}_A\otimes\mathbbm{I}_E)\rho_{BAEX}(U^{\dag}_{BA}\otimes\mathbbm{I}_A\otimes\mathbbm{I}_E)$. Then the auxiliary interacts with the environment through the unitary operator $U_{AE}$ the state of the composite system 
is given by $\rho^2_{BAEX}=(\mathbbm{I}_B\otimes U_{AE}\otimes\mathbbm{I}_X)\rho^1_{BAEX}(\mathbbm{I}_B\otimes U^{\dag}_{AE}\otimes\mathbbm{I}_X)$. Now, to implement NPOVM type-2 we perform a three-body projective measurement on the auxiliary, environment, and external system, then trace out the environment and external system. Therefore, after implementation of the projective measurement on the auxiliary environment and auxiliary system, the state of the composite system is given by $\rho^2_{BAEX}=(\mathbbm{I}_B\otimes \left(U_M^{N^2}\right)\mathcal{A}''_i\left(U_M^{N^2}\right)^\dagger)(\mathbbm{I}_B\otimes U^{N^2}_M)(\mathbbm{I}_B\otimes U_{AE}\otimes\mathbbm{I}_X)\rho^1_{BAEX}(\mathbbm{I}_B\otimes U^{\dag}_{AE}\otimes\mathbbm{I}_X)(\mathbbm{I}_B\otimes \left(U_M^{N^2}\right)\mathcal{A}''_i\left(U_M^{N^2}\right)^\dagger)$.  Here, $\{\left(U_M^{N^2}\right)\mathcal{A}''_i\left(U_M^{N^2}\right)^\dagger)\}$ is the measurement setting. The distillable energy by NPOVM type-1 operation is given by

\begin{eqnarray}\label{eq7}
S^{NP_2}_{D}
=\max_{U_M^{N^2}}~ \sum_{i\in\mathcal{I}} p''_i \,
   \text{tr}\!\Big[ Z'' \big(\mathbbm{I}_B \otimes \left(U_M^{N^2}\right) (U_{AE}^{\dag}\otimes\mathbbm{I}_X) \mathcal{A}''_i \mathcal{A}''^{\dag}_i \left(U_M^{N^2}\right)^{\dag} (U_{AE}\otimes\mathbbm{I}_X)\big)
   \rho_{BAEX} \Big] \nonumber.
\end{eqnarray}

{Here, we may also define $\widetilde{U}_{AEX} = U_M^{N^2}(U_{AE} \otimes \mathbbm{I}_X)$. Similar to the NPOVM type-1 case, the optimization over the redefined unitary $\widetilde{U}_{AEX}$ eliminates the effect of noise. Following the same reasoning, it can be shown that 
the maximum stochastically extractable energy and the average extractable energy under an NPOVM type-2 operation also become independent of both the type and the strength of the noise.
{Using a similar line of reasoning, one can also demonstrate that the average energy remains independent of the noise strength.}
}

\section{Set of considered POVMs is not a subset of the set of considered type-1 NPOVMs}\label{aD}

{Since all the examples discussed in Sec.~\ref{sec4} prove NPOVMs of type-1 are always equally or more advantageous than POVMs in the context of energy extraction, one might wonder if the set of POVMs applied on the auxiliary is a subset of the NPOVMs of type-1 applied on the same. Here by the set of type-1 NPOVMs, we basically mean the entire general set over which the optimization is performed, which may also contain POVM elements. Here, we demonstrate that this is not the case; specifically, we will show that the set of all POVMs is not contained within the set of type-1 NPOVMs.}

{In the considered scenario, we took the initial state of the total system as $\ket{e}_B\bra{e} \otimes \ket{e}_A\bra{e}\otimes \ket{g}_E\bra{g} \otimes \tr_{BA}[\rho_{BAE}^2]$. Each of  $\ket{e}_B\bra{e}$, $\ket{e}_A\bra{e}$ and $\ket{g}_E\bra{g}$ is a rank-one state, and the dimension of each of the smallest subsystems is considered to be two. Initially, the battery ($B$) and auxiliary ($A$) undergo a unitary evolution, which, in general, creates entanglement between $B$ and $A$. As a result, the rank of the local state of $A$ increases. Let us consider the eigenvalues of the state of $A$ at the end of its interaction with $B$ to be $\lambda_1$ and $\lambda_2$. Since, at this instant, $A$ and the $E$ (i.e., the environment) are in a product state and $E$ is in a pure state, the set of eigenvalues of the composite system $AE$ is given by, $\{\lambda_1,\lambda_2,0,0\}$. After the evolution of $BA$, noise acts on $A$ through an interaction between $A$ and $E$. At the end of the interaction, the state of $BAE$ is $\rho_{BAE}^2$.} {As the eigenspectrum of an operator remains invariant under unitary operation, the set of eigenvalues of the joint state of $A$ and $E$ will remain unchanged in this evolution.
Hence the eigenvalues of $\rho_{AE}^2=\tr_{B}(\rho^2_{BAE})$ are $\{\lambda_1,\lambda_2,0,0\}$.  Let the set of eigenvalues of $\rho_A^2=\tr_{BE}(\rho_{BAE}^2)$ and  $\rho^2_E=\tr_{BA}(\rho_{BAE}^2)$ be $\{x_1,x_2\}$ and $\{y_1,y_2\}$, respectively. As the initial state, $\rho_X$, of $X$ is taken same as the marginal state of $E$, i.e., $\rho_X=\rho^2_E$, and $X$ is not entangled with $BAE$, the eigenvalues of $AX$ will be $\{\ x_1 y_1,x_1 y_2,x_2y_1,x_2 y_2\}$ which, in general, is different from the eigenvalues of $AE$.
At this moment, we apply the POVM and type-1 NPOVM on $A$ by performing a joint projective measurement on $AX$ and $AE$, respectively.}

{A rank-one projective measurement in an arbitrary direction on a system can be implemented by operating a unitary on the system and measuring it in the computational basis. Therefore, performing the POVM and NPOVM of type-1 on the auxiliary are equivalent to measuring the states $\widetilde{U}_{AX}\rho_A^2\otimes\rho_X \widetilde{U}_{AX}^\dagger$ and $\widetilde{U}_{AE} \rho_{AE}^2 \widetilde{U}_{AE}^\dagger$ in the computational basis, where $\widetilde{U}_{AX}$ and $\widetilde{U}_{AE}$ are, respectively, the unitaries acting on $AX$ and $AE$, that specify 
the direction of the corresponding projective measurements. We can even consider $\widetilde{U}_{AX}=\mathbb{I}_A\otimes\mathbb{I}_X$ as an example of a POVM, which represents measuring $AX$ in the computational basis. For the POVM to be a subset of the set of NPOVMs, this simple example should also be an element of the set of NPOVMs. In that case, there should exist a $\widetilde{U}_{AE}=\widetilde{U}_{\text{ex}}$ such that, $\widetilde{U}_{\text{ex}} \rho_{AE}^2 \widetilde{U}_{\text{ex}}^\dagger=\rho_A^2 \otimes \rho^2_X$. Since the eigenspectrum of $\rho^2_{AE}$ is $\{\lambda_1,\lambda_2,0,0\}$, while that of $\rho_A^2\otimes \rho_X$ is $\{x_1y_1,x_1y_2, x_2y_1,x_2y_2\}$, and unitaries cannot change eigenvalues, for the condition, $\widetilde{U}_{\text{ex}} \rho_{AE}^2 \widetilde{U}_{\text{ex}}^\dagger=\rho_A^2 \otimes \rho_X$, to hold, we must have  
$\{\lambda_1,\lambda_2,0,0\}\equiv\{x_1y_1,x_1y_2, x_2y_1,x_2y_2\}$, which implies either $\{x_1,x_2\} \equiv \{\lambda_1, \lambda_2\}$ and $\{y_1,y_2\}\equiv\{0,1\}$ or $\{y_1,y_2\} \equiv \{\lambda_1, \lambda_2\}$ and $\{x_1,x_2\}\equiv\{0,1\}$. But this is not true in general, and therefore we conclude that the set of all type-1 NPOVMs does not form a superset of the set of all POVMs.}

\label{sec-Appendix}
\appendix

\twocolumngrid

\bibliography{Battery}

\end{document}